\documentclass[11pt,english,vecarrow]{extarticle}
\usepackage[T1]{fontenc}
\usepackage[utf8]{inputenc}
\usepackage[letterpaper]{geometry}
\usepackage{babel}
\usepackage{float}
\usepackage{amsmath}
\usepackage{amssymb}
\usepackage{stmaryrd}
\usepackage{graphicx}
\usepackage{setspace}
\usepackage{wasysym}

\makeatletter
\numberwithin{equation}{section}
\numberwithin{figure}{section}
\numberwithin{table}{section}

\usepackage{bbm}
\usepackage{tensind}
\tensordelimiter{?}
\usepackage{multicol}
\usepackage[labelfont={sc,small},textfont=small]{caption}
 
\newcommand{\ket}[1]{\rvert #1 \rangle} 
\newcommand{\braket}[2]{\left \langle #1 \middle \rvert #2 \right \rangle}
\newcommand{\braxket}[3]{\left \langle #1 \middle \rvert #2 \middle \rvert #3 \right \rangle}

\makeatother

\begin{document}
\title{From quantum electrodynamics to a geometric gauge theory of classical
electromagnetism\\
\quad{}}
\author{\doublespacing{}\textsc{Adam Marsh}\thanks{\textsc{Electronic address: adammarsh@berkeley.edu}}}
\date{\ }
\maketitle
\begin{abstract}
A relativistic version of the correspondence principle, a limit in
which classical electrodynamics may be derived from QED, has never
been clear, especially when including gravitational mass. Here we
introduce a novel classical field theory formulation of electromagnetism,
and then show that it approximates QED in the limit of a quantum state
which corresponds to a classical charged continua. Our formulation
of electromagnetism features a Lagrangian which is gauge invariant,
includes a classical complex field from which a divergenceless four-current
may be derived, and reproduces all aspects of the classical theory
of charged massive continua without any quantum effects. Taking a
geometric approach, we identify the four-current as being in the direction
of extremal phase velocity of the classical field; the field equations
of motion determine this phase velocity as being equal to the mass,
which makes the rest density proportional to the squared modulus of
the field. 
\end{abstract}
\tableofcontents{}

\section{Introduction}

\subsection{Motivation}

The (quantized, minimally coupled) Dirac equation describes (relativistic,
quantum mechanical) electrons (and positrons) interacting electromagnetically.
The Dirac spinor matter field comprises four complex components smoothly
defined at each point of spacetime; its values as operators provide
a description of multiple particles via quantum electrodynamics (QED).
The Dirac equation may be derived from the Dirac Lagrangian, which
is invariant under $U(1)$ gauge transformations.

Maxwell's equations describe (relativistic, classical) charged massive
continua interacting electromagnetically. The four-current is a divergenceless
four-vector field on spacetime whose direction is that of the continuum
of particles at that point, and whose length is the number of particles
per unit space-like volume orthogonal to that direction. Maxwell's
equations are usually derived from a Lagrangian which includes this
four-current, does not include a matter field, and is not invariant
under $U(1)$ gauge transformations.

While the classical electromagnetic gauge potential equation of motion
may be obtained from QED by the stationary phase approximation, there
is no way to obtain a classical matter field yielding a classical
four-current using this approximation, since the scalars obtained
must anti-commute (i.e. they are Grassmann numbers). 

Our aim in this paper is to construct an alternative Lagrangian for
classical charged massive continua interacting electromagnetically
which (1) is based upon a classical matter field whose equations of
motion yield a classical four-current, (2) is invariant under gauge
transformations, and (3) may be obtained as a limit of quantum electrodynamics.
An additional goal is a detailed presentation of this alternative
theory in terms of both real and complex geometry. 

\subsection{Geometry}

The evolution of gauge theory was from its beginning geometric and
tied to electromagnetism. Weyl \cite{Weyl 1918} first introduced
the concept of gauge invariance as local spacetime scale invariance
(thus the name) in an attempt to unify general relativity with electromagnetism,
later repurposing it more successfully \cite{Weyl 1929} as an extension
of the global phase invariance of the wave function in Dirac theory
to local phase invariance, which yields Maxwell's equations. The resulting
``gauge principle,'' formulated as a procedure for extending global
symmetries to local symmetries, was then generalized to higher dimensional
complex ``rotations'' by Yang and Mills \cite{YangMills}. 

This set of ideas was eventually given an even more geometric formulation
in the language of fiber bundles (see e.g. \cite{Bleecker}), which
also accounts for global considerations. A scalar matter field (a
generic term to encompass both scalar and operator field values) is
defined as a section of a complex vector bundle over a (possibly curved)
spacetime manifold; in analogy with tangent vector fields, the gauge
potential is a connection defining parallel transport of these vectors,
the field strength is the curvature of this connection, and a gauge
transformation is a change of the frame defining the matter field
components.

In the viewpoint adopted here, this geometric formulation positions
the matter field as the primary quantity, with the gauge potential
defining its parallel transport. Classical electromagnetism however,
despite being the prototypical gauge theory, is typically described
as a gauge theory with the four-current inserted by hand as an external
quantity, while the notions of a gauge potential and field strength
are preserved even in the absence of an associated matter field. Moreover,
the Lagrangian is not gauge invariant, which from a geometric point
of view is nonsensical. We will therefore refer to this formulation
absent a matter field as ``quasi-gauge electromagnetism.''

\subsection{Overview}

We would like to formulate the classical theory of charged massive
continua as a gauge theory including a matter field. This matter field
should determine a divergenceless four-current; but unlike in quantum
theory, its equations of motion should not result in solutions with
quantum characteristics. In particular, we should have no need to
interpret the ``on-shell'' matter fields which satisfy these equations
in terms of particles or probabilities; instead, they should determine
a general classical four-current.

In the following we provide a geometric description of such a theory,
which is based upon a gauge-invariant Lagrangian which includes a
matter field, and which, like quasi-gauge electromagnetism, reproduces
the full classical theory of charged massive continua, without any
quantum characteristics. In addition, QED is shown to simplify to
this theory under certain conditions. We will refer to this theory
as ``matter field electromagnetism.''

In Section \ref{sec:Geometric-U1-gauge-theory} we summarize our geometric
view of gauge theory and describe various geometrical quantities associated
with our $U(1)$ gauge theory from both the real and complex viewpoints,
taking pains to construct a consistent picture in the real case. Section
\ref{sec:Matter-field-electromagnetism} defines our Lagrangian for
matter field electromagnetism, in terms of both geometry and complex
algebra, and derives the equations of motion along with the results
of Noether's theorem. Section \ref{sec:From-QED-to-MFEM} constructs
a quantum state corresponding to a classical current and shows that,
as desired, QED in the limit of such a state results in the matter
field and equations of motion from matter field electromagnetism. 

Throughout the paper we will use natural units, where the constants
$c=G=\hbar=1$, and the mostly pluses spacetime metric signature,
where in an orthonormal frame the metric is $g_{\mu\nu}=\mathrm{diag}\left(-1,1,1,1\right)$.

\section{\label{sec:Geometric-U1-gauge-theory}Geometric $U(1)$ gauge theory}

\subsection{A geometric view of gauge theory}

A common way to describe a scalar gauge theory is in terms of a multiplet
of (usually complex) matter fields $\Phi^{a}$ and a Lagrangian which
is written in terms of these fields along with their coordinate derivatives
$\partial_{\mu}\Phi^{a}$. The Lagrangian is then noted to be invariant
under a global gauge transformation, a complex matrix transformation
of the $\Phi^{a}$ treated as vector components, where the matrix
is an element of a matrix group, usually $U(1)$ or $SU(n)$ (hereafter
referred to as simply $SU(n)$). The principle of minimal coupling
then prescribes the introduction of a gauge potential to replace coordinate
derivatives with gauge covariant derivatives. This promotes the global
gauge invariance to a local gauge invariance, wherein the Lagrangian
is invariant under multiplication by the matter field at each point
by a different but smoothly varying matrix in $SU(n)$.

The geometric view of scalar gauge theory we take here (detailed in
\cite{Marsh}) is based upon a vector bundle $(E,M,X)$ over spacetime
$M$, with the fiber $X\cong\mathbb{C}^{n}$ called the internal space.
A matter field is a section of $E$, or equivalently an $X$-valued
0-form (function) on $M$. A choice of gauge in a region of $M$ is
a frame, a choice of orthonormal basis for the fiber $X_{p}$ at each
point $p$ in the region, which allows us to express the matter field
as a gauge-dependent $\mathbb{C}^{n}$-valued 0-form we denote $\vec{\Phi}$,
with components $\Phi^{a}$ smoothly defined at each point. A gauge
transformation is a change of frame, which in a choice of gauge corresponds
to a matrix element smoothly defined at each point which we denote
$\check{\gamma}^{-1}\in SU\left(n\right)$; the check decoration indicates
a matrix value, and the element is defined as an inverse matrix applied
to the basis so that the components $\Phi^{a}$ at each point in the
region transform as $\vec{\Phi}\rightarrow\check{\gamma}\vec{\Phi}$.
A gauge transformation is thus a ``rotation'' of the internal space
basis at each point, or more precisely a linear transformation which
leaves the complex inner product on the fiber $X$ invariant, along
with the complex volume element $\mathsf{e}_{1}\wedge\cdots\wedge\mathsf{e}_{n}$
for $n>1$; here $\mathsf{e}_{a}$ is a complex orthonormal basis
of $X$, and we use a sans serif font for internal space basis vectors
to distinguish them from spacetime tangent space basis vectors.

In analogy with the tangent bundle, we introduce a connection defining
parallel transport, an $su(n)$-valued 1-form we denote 
\begin{equation}
\check{\Gamma}\equiv-iq\check{A}\equiv-iqA^{a}{}_{b\mu},
\end{equation}
where the gauge potential $A^{a}{}_{b\mu}$ is thus a hermitian matrix-valued
1-form with a Greek spacetime index. Geometrically, $-iqA^{a}{}_{b\mu}$
is the $a^{\mathrm{th}}$ component of the difference between the
frame $\mathsf{e}_{b}$ and its parallel transport in the direction
$e_{\mu}$. 

The gauge covariant derivative, which is geometrically the infinitesimal
difference between $\vec{\Phi}$ and its parallel transport, can then
be written in terms of forms and components as
\begin{equation}
\begin{aligned}\mathrm{D}\vec{\Phi} & =\mathrm{d}\vec{\Phi}+\check{\Gamma}\vec{\Phi},\\
\mathrm{D}_{\mu}\Phi^{a} & =\partial_{\mu}\Phi^{a}-iqA^{a}{}_{b\mu}\Phi^{b}.
\end{aligned}
\end{equation}
The connection defines a $su(n)$-valued curvature 2-form written
in terms of the field strength $\check{F}$ as $-iqF^{a}{}_{b\mu\nu}$,
where 
\begin{equation}
\begin{aligned}\check{F} & \equiv\mathrm{d}\check{A}-iq\check{A}\wedge\check{A},\\
F^{a}{}_{b\mu\nu} & =\partial_{\mu}A^{a}{}_{b\nu}-\partial_{\nu}A^{a}{}_{b\mu}-iq\left(A^{a}{}_{c\mu}A^{c}{}_{b\nu}-A^{a}{}_{c\nu}A^{c}{}_{b\mu}\right).
\end{aligned}
\end{equation}
Under a gauge transformation $\check{\gamma}^{-1}$, these quantities
transform as follows:
\begin{equation}
\begin{aligned}\vec{\Phi} & \rightarrow\check{\gamma}\vec{\Phi}\\
\check{F} & \rightarrow\check{\gamma}\check{F}\check{\gamma}^{-1}\\
\check{A} & \rightarrow\check{\gamma}\check{A}\check{\gamma}^{-1}+\frac{i}{q}\check{\gamma}\mathrm{d}\check{\gamma}^{-1}
\end{aligned}
\end{equation}

In the case of $U(1)$ gauge theory, the matter field is complex-valued
and denoted $\Phi$, while the gauge potential $A_{\mu}$ is real-valued.
A gauge transformation is then usually written $\gamma^{-1}\equiv e^{-iq\Lambda}$,
under which the above relations simplify to:
\begin{equation}
\begin{aligned}\mathrm{D}_{\mu}\Phi & =\partial_{\mu}\Phi-iqA{}_{\mu}\Phi\\
F_{\mu\nu} & =\partial_{\mu}A{}_{\nu}-\partial_{\nu}A{}_{\mu}\\
\Phi & \rightarrow e^{iq\Lambda}\Phi\\
F_{\mu\nu} & \rightarrow F_{\mu\nu}\\
A_{\mu} & \rightarrow A_{\mu}+\partial_{\mu}\Lambda
\end{aligned}
\end{equation}
\begin{figure}[H]
\noindent \begin{centering}
\includegraphics[scale=1.15]{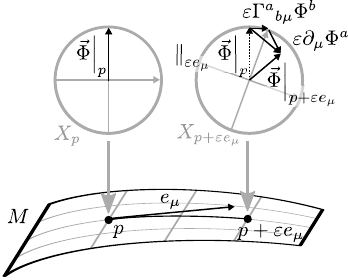}
\par\end{centering}
\caption{A gauge theory is a bundle over spacetime $M$ with fiber a vector
space $X$, and matter field $\vec{\Phi}$ a section of this bundle.
If we choose a gauge, an orthonormal basis for each $X_{p}$, the
matter field can be written in terms of components $\Phi^{a}$, with
a gauge transformation being a smoothly defined change of orthonormal
basis for each $X_{p}$. The gauge covariant derivative in a given
gauge can then be written $\mathrm{D}_{\mu}\Phi^{a}=\partial_{\mu}\Phi^{a}+\Gamma^{a}{}_{b\mu}\Phi^{b}=\partial_{\mu}\Phi^{a}-iqA^{a}{}_{b\mu}\Phi^{b}$.
Geometrically, $\varepsilon\mathrm{D}_{\mu}\Phi^{a}=\left.\vec{\Phi}\right|_{p+\varepsilon e_{\mu}}-\parallel_{\varepsilon e_{\mu}}\left.\vec{\Phi}\right|_{p}$
is the difference between the matter field $\left.\vec{\Phi}\right|_{p+\varepsilon e_{\mu}}$
at a point infinitesimally displaced in the direction $e_{\mu}$ from
$p$ and its parallel transport from $p$, which we denote $\parallel_{\varepsilon e_{\mu}}\left.\vec{\Phi}\right|_{p}$.}
\end{figure}

\subsection{The gauge potential}

In $U(1)$ gauge theory, the matter field value at each point in spacetime
is a complex number $\Phi\in\mathbb{C}$, which from our geometric
viewpoint is the single complex component of an intrinsic complex
vector in a given gauge (choice of unit length complex vector). A
gauge transformation (new choice of unit length complex vector) $\Phi\rightarrow e^{iq\Lambda}\Phi$,
or infinitesimally $\Phi\rightarrow\Phi+iq\Lambda\Phi$, changes the
complex component, but does not change the intrinsic complex vector.
Note that a choice of unit length complex vector is also a choice
of volume element, which gauge transformations are only allowed to
alter in one complex dimension; i.e. in one dimension we consider
$U(1)$ instead of $SU(1)$. 

Another unique feature of $U(1)$ gauge theory is that $U(1)\cong SO(2)$,
the complete group of rotations in the decomplexified space, instead
of a subgroup $SU(n)\subset SO(2n)$ of these rotations. The matter
field at each point may therefore be viewed as a real vector $\vec{\Phi}\in\mathbb{R}^{2}$,
with a gauge transformation a new choice of real orthonormal basis
for the internal space which leaves $\vec{\Phi}$ unchanged but transforms
its components $\Phi^{a}$. We will adopt this view as our primary
one, and will take it quite literally; but we will utilize the complex
notation in parallel with the vector notation, since electromagnetism
is almost universally expressed this way, and the definition of the
gauge potential $A_{\mu}$ is dependent upon the complex viewpoint.

We associate the rotation of a real vector $\vec{\Phi}\in\mathbb{R}^{2}$
by $\theta$ radians with the complex multiplication $e^{i\theta}\Phi$
in $\mathbb{C}$. This rotation is usually depicted as being in the
counterclockwise direction, since the real axis is usually depicted
as the horizontal axis. Therefore, going forward we will use the word
``rotation'' to mean ``counterclockwise rotation in this depiction''
or ``rotation from the positive real towards the positive imaginary
axis.''

Now, $-iqA_{\mu}$ is the imaginary number which when multiplied by
a unit basis vector gives the infinitesimal difference between the
basis vector at a displaced point and its parallel transport. Since
this displacement is infinitesimal and rotates a unit vector, the
rotation is by $-qA_{\mu}$ radians, so that we can view $q$ as a
conversion factor between units, i.e. a rotational multiplier specifying
internal space radians per unit spacetime length for unit $A_{\mu}$.
We may then write the (counterclockwise instantaneous) angular velocity
(relative to parallel transport) of the internal space frame (gauge)
in the $e_{\mu}$ direction as 
\begin{align}
A_{\mu}^{\varangle} & \equiv-qA_{\mu}
\end{align}
in radians per coordinate length; a positive gauge potential value
then corresponds to a clockwise frame rotation if $q$ is positive.
\begin{figure}[H]
\noindent \begin{centering}
\includegraphics[scale=1.15]{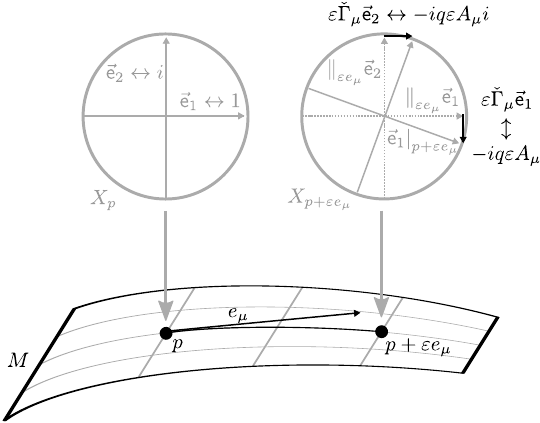}
\par\end{centering}
\caption{In a given choice of gauge, the $so(2)$-valued connection $\check{\Gamma}$
is a matrix which applied to any frame vector $\vec{\mathsf{e}}_{a}$
yields the difference between the frame vector $\left.\vec{\mathsf{e}}_{a}\right|_{p+\varepsilon e_{\mu}}$
at the infinitesimally displaced point and the parallel transported
frame vector $\parallel_{\varepsilon e_{\mu}}\vec{\mathsf{e}}_{a}$.
In the complex view, the connection is $-iqA$, a complex number which
when multiplied by any frame vector yields the difference between
the frame vector at the new point and the parallel transported frame
vector, again as a complex number. For example, the internal space
frame vector $\vec{\mathsf{e}}_{2}$ corresponds to $i$, so that
the difference is $-iqAi=qA$, which is a positive real number if
$q$ and $A$ are. $q$ may therefore be viewed as a conversion factor
which in the chosen units makes $qA_{\mu}$ the clockwise angular
velocity of the frame (relative to parallel transport) in the $e_{\mu}$
direction. Note that if $\mu\protect\neq0$ this \textquotedblleft velocity\textquotedblright{}
will be in radians per unit distance instead of radians per unit time.}
\end{figure}

\subsection{\label{subsec:Inner-products}Inner products}

The frame rotations of the previous section imply that we define the
real inner product on the internal space $\mathbb{R}^{2}\cong\mathbb{C}$
as the real part of the complex inner product $\left\langle \vec{\Phi},\vec{\Psi}\right\rangle _{\mathbb{R}}=\mathrm{Re}\left\langle \Phi,\Psi\right\rangle _{\mathbb{C}}$;
in particular, the squared length of a matter field vector may then
be written
\begin{equation}
\begin{aligned}\left\langle \vec{\Phi},\vec{\Phi}\right\rangle _{\mathbb{R}} & =\left\Vert \vec{\Phi}\right\Vert ^{2}=\Phi^{a}\Phi_{a}\\
=\left\langle \Phi,\Phi\right\rangle _{\mathbb{C}} & =\Phi^{*}\Phi=\left|\Phi\right|^{2},
\end{aligned}
\end{equation}
where $\left|\Phi\right|$ is the modulus and $\Phi^{*}$ is the complex
conjugate of the complex number $\Phi$. For the field strength, we
have
\begin{equation}
\begin{aligned}\left\langle F,F\right\rangle  & =\frac{1}{2}F_{\mu\nu}F^{\mu\nu}\\
 & =\frac{1}{2}\left(\partial_{\mu}A_{\nu}-\partial_{\nu}A_{\mu}\right)\left(\partial^{\mu}A^{\nu}-\partial^{\nu}A^{\mu}\right)\\
 & =\partial_{\mu}A_{\nu}\partial^{\mu}A^{\nu}-\partial_{\mu}A_{\nu}\partial^{\nu}A^{\mu},
\end{aligned}
\label{eq:FF}
\end{equation}
where we recall the $k$-form inner product relation $\left\langle \varphi,\psi\right\rangle _{\textrm{form}}=\frac{1}{k!}\varphi_{\mu_{1}\dots\mu_{k}}\psi^{\mu_{1}\dots\mu_{k}}$
and in the last line we swap dummy indices and combine terms.

We can write the ``squared length'' of the gauge covariant derivative
as 
\begin{equation}
\begin{aligned}\left\langle \mathrm{D}_{\mu}\vec{\Phi},\mathrm{D}_{\mu}\vec{\Phi}\right\rangle  & \equiv\left(\mathrm{D}^{\mu}\Phi\right)^{*}\mathrm{D}_{\mu}\Phi\\
 & =\left(\partial^{\mu}\Phi^{*}+iqA^{\mu}\Phi^{*}\right)\left(\partial_{\mu}\Phi-iqA_{\mu}\Phi\right)\\
 & =\partial^{\mu}\Phi^{*}\partial_{\mu}\Phi+iqA^{\mu}\left(\Phi^{*}\partial_{\mu}\Phi-\Phi\partial_{\mu}\Phi^{*}\right)+q^{2}\left|\Phi\right|^{2}A_{\mu}A^{\mu}.
\end{aligned}
\end{equation}
If we define 
\begin{equation}
\begin{aligned}\Phi & \equiv\left|\Phi\right|e^{i\varphi}\\
\Rightarrow\mathrm{D}_{\mu}\Phi & =\frac{\Phi}{\left|\Phi\right|}\mathrm{D}_{\mu}\left|\Phi\right|+i\Phi\mathrm{D}_{\mu}\varphi\\
 & =\partial_{\mu}\Phi-iqA_{\mu}\Phi\\
 & =\frac{\Phi}{\left|\Phi\right|}\partial_{\mu}\left|\Phi\right|+i\Phi\partial_{\mu}\varphi-iqA_{\mu}\Phi,
\end{aligned}
\label{eq:DPhi-components}
\end{equation}
we see that geometrically 
\begin{equation}
\begin{aligned}\mathrm{D}_{\mu}\left|\Phi\right| & =\partial_{\mu}\left|\Phi\right|,\\
\mathrm{D}_{\mu}\varphi & =\partial_{\mu}\varphi-qA_{\mu}
\end{aligned}
\end{equation}
are the difference in length and angle between the matter field and
its parallel transport per unit distance in the direction $e_{\mu}$.
In terms of these quantities, we have 
\begin{equation}
\begin{aligned}\left\langle \mathrm{D}_{\mu}\Phi,\mathrm{D}_{\mu}\Phi\right\rangle  & =\left(\frac{\Phi^{*}}{\left|\Phi\right|}\mathrm{D}^{\mu}\left|\Phi\right|-i\Phi^{*}\mathrm{D}^{\mu}\varphi\right)\left(\frac{\Phi}{\left|\Phi\right|}\mathrm{D}_{\mu}\left|\Phi\right|+i\Phi\mathrm{D}_{\mu}\varphi\right)\\
 & =\left\langle \mathrm{D}_{\mu}\left|\Phi\right|,\mathrm{D}_{\mu}\left|\Phi\right|\right\rangle +\left|\Phi\right|^{2}\left\langle \mathrm{D}_{\mu}\varphi,\mathrm{D}_{\mu}\varphi\right\rangle .
\end{aligned}
\label{eq:DPhiDPhi-complex-components}
\end{equation}

\subsection{\label{subsec:Spacetime-gradient}The spacetime gradient}

\begin{figure}[H]
\noindent \begin{centering}
\includegraphics[scale=1.15]{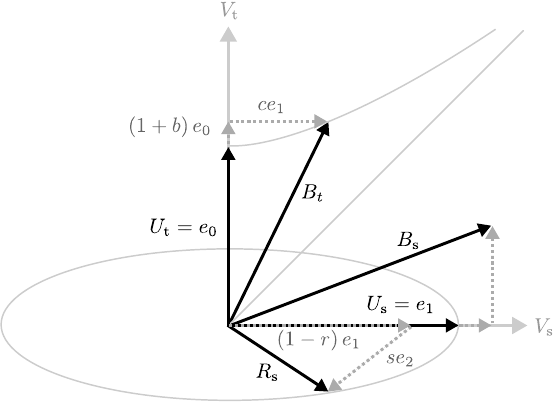}
\par\end{centering}
\caption{\label{fig:Extremal-direction}For a time-like vector $V_{\mathrm{t}}$
and a space-like vector $V_{\mathrm{s}}$, the unit vector $U$ parallel
to $V$ is the unit vector for which $\left|V_{\mu}U^{\mu}\right|$
is a minimum compared to any boosted unit vector $B$, and is a maximum
compared to any rotated unit vector $R$. }
\end{figure}

$\left\langle \mathrm{D}_{\mu}\varphi,\mathrm{D}_{\mu}\varphi\right\rangle $
is the magnitude of the four-vector $\mathrm{D}^{\mu}\varphi$, which
is a ``spacetime gradient.'' Since we find no ready reference, we
characterize the direction of this four-vector here. For any non-null
spacetime four-vector $V^{\mu}$, we may choose an orthonormal basis
for the tangent space for which $V=Le_{\parallel}$, with $\left\langle e_{\parallel},e_{\parallel}\right\rangle =\pm1$
and any other basis four-vector $e_{\perp}$ orthogonal to $V$. For
any unit four-vector $\left\langle B,B\right\rangle =\pm1$ which
is a boost of $e_{\parallel}$ in the same part of the light cone,
we have
\begin{equation}
\begin{aligned}B & =e_{\parallel}+be_{\parallel}+ce_{\perp}\\
\Rightarrow V_{\mu}B^{\mu} & =\pm L\left(1+b\right)
\end{aligned}
\end{equation}
for positive numbers $b$ and $c$ (see Figure \ref{fig:Extremal-direction});
$e_{\parallel}$ is therefore the direction $U$ for which the absolute
value $\left|V_{\mu}U^{\mu}\right|$ is at a minimum under boosts.
Similarly, for any unit four-vector $\left\langle R,R\right\rangle =\pm1$
which is a rotation of $e_{\parallel}$ in the same part of the light
cone (excluding $-e_{\parallel}$), we have
\begin{equation}
\begin{aligned}R & =e_{\parallel}-re_{\parallel}+se_{\perp}\\
\Rightarrow V_{\mu}R^{\mu} & =\pm L\left(1-r\right)
\end{aligned}
\end{equation}
for positive numbers $r<2$ and $s$; $e_{\parallel}$ is therefore
the direction $U$ for which the absolute value $\left|V_{\mu}U^{\mu}\right|$
is at a maximum under rotations. 

With these results in hand, we may geometrically describe $\mathrm{D}^{\mu}\varphi$
as pointing in the direction $U^{\mu}$ in which $\left|\mathrm{D}_{U}\varphi\right|=\left|U^{\mu}\mathrm{D}_{\mu}\varphi\right|$
is smallest under boosts and largest under rotations (the latter corresponding
to the usual gradient description as the ``direction of steepest
ascent''). In the interest of brevity we may say that $\mathrm{D}^{\mu}\varphi$
points in the direction $U$ in which $\mathrm{D}_{U}\varphi$ is
extremal; in particular, $\mathrm{D}_{U^{\perp}}\varphi=0$ for any
$U^{\perp}$ orthogonal to $U$. We may then say that $\left|\left\langle \mathrm{D}_{\mu}\varphi,\mathrm{D}_{\mu}\varphi\right\rangle \right|$
is the squared angular difference between the matter field and its
parallel transport per unit distance in the direction in which this
difference is extremal.

Note that $\mathrm{D}_{\mu}\Phi$ has values which are complex and
therefore not ordered, disallowing the above interpretation, and there
will in general be different directions in which the change in the
modulus and phase are extremal.

\subsection{\label{subsec:Matter-field-rotation}Matter field rotation}

The imaginary part of the complex inner product is just the real inner
product with one of the arguments rotated by $\pi/2$, i.e. 
\[
\begin{aligned}\mathrm{Im}\left\langle \Psi,\Phi\right\rangle _{\mathbb{C}} & =\mathrm{Re}\left(-i\left\langle \Psi,\Phi\right\rangle _{\mathbb{C}}\right)\\
 & =\left\langle \vec{\Psi},-\overrightarrow{\left(i\Phi\right)}\right\rangle _{\mathbb{R}}\\
 & =\left\langle \overrightarrow{\left(i\Psi\right)},\vec{\Phi}\right\rangle _{\mathbb{R}}.
\end{aligned}
\]
Since the real inner product is geometrically a projection, we may
express the lengths of the components of $\vec{\Psi}$ parallel and
perpendicular to $\vec{\Phi}$ as
\begin{equation}
\begin{aligned}\vec{\Psi}^{\parallel\vec{\Phi}} & =\frac{\left\langle \vec{\Psi},\vec{\Phi}\right\rangle _{\mathbb{R}}}{\left\Vert \vec{\Phi}\right\Vert }=\frac{\mathrm{Re}\left\langle \Psi,\Phi\right\rangle _{\mathbb{C}}}{\left|\Phi\right|}\\
 & =\frac{\left(\Phi^{*}\Psi+\Phi\Psi^{*}\right)}{2\left|\Phi\right|},\\
\vec{\Psi}^{\perp\vec{\Phi}} & =\frac{\left\langle \vec{\Psi},\overrightarrow{\left(i\Phi\right)}\right\rangle _{\mathbb{R}}}{\left\Vert \vec{\Phi}\right\Vert }=\frac{-\mathrm{Im}\left\langle \Psi,\Phi\right\rangle _{\mathbb{C}}}{\left|\Phi\right|}\\
 & =\frac{-i\left(\Phi^{*}\Psi-\Phi\Psi^{*}\right)}{2\left|\Phi\right|}.
\end{aligned}
\end{equation}
In particular, defining 
\[
\begin{aligned}\mathrm{D}_{\mu}^{\parallel}\vec{\Phi} & \equiv\left(\mathrm{D}_{\mu}\vec{\Phi}\right)^{\parallel\vec{\Phi}}\\
\mathrm{D}_{\mu}^{\perp}\vec{\Phi} & \equiv\left(\mathrm{D}_{\mu}\vec{\Phi}\right)^{\perp\vec{\Phi}},
\end{aligned}
\]
we have
\begin{equation}
\begin{aligned}2\left\Vert \vec{\Phi}\right\Vert \mathrm{D}_{\mu}^{\parallel}\vec{\Phi} & =2\left\langle \mathrm{D}_{\mu}\vec{\Phi},\vec{\Phi}\right\rangle _{\mathbb{R}}\\
 & =2\mathrm{Re}\left\langle \mathrm{D}_{\mu}\Phi,\Phi\right\rangle _{\mathbb{C}}\\
 & =\Phi^{*}\mathrm{D}_{\mu}\Phi+\Phi\left(\mathrm{D}_{\mu}\Phi\right)^{*}\\
 & =\Phi^{*}\partial_{\mu}\Phi+\Phi\partial_{\mu}\Phi^{*}\\
 & =\partial_{\mu}\left|\Phi\right|^{2},\\
2\left\Vert \vec{\Phi}\right\Vert \mathrm{D}_{\mu}^{\perp}\vec{\Phi} & =2\left\langle \mathrm{D}_{\mu}\vec{\Phi},\overrightarrow{\left(i\Phi\right)}\right\rangle _{\mathbb{R}}\\
 & =2\mathrm{Im}\left\langle \Phi,\mathrm{D}_{\mu}\Phi\right\rangle _{\mathbb{C}}\\
 & =-i\left(\Phi^{*}\mathrm{D}_{\mu}\Phi-\Phi\left(\mathrm{D}_{\mu}\Phi\right)^{*}\right)\\
 & =-i\left(\Phi^{*}\partial_{\mu}\Phi-\Phi\partial_{\mu}\Phi^{*}-2iq\left|\Phi\right|^{2}A_{\mu}\right).
\end{aligned}
\label{eq:DPhi-component-expressions}
\end{equation}
As expected, since it measures the change in the length of $\vec{\Phi}$,
we can see that $\mathrm{D}_{\mu}^{\parallel}\vec{\Phi}$ is independent
of $A_{\mu}$, which affects only rotations; it may also be written
\begin{equation}
\begin{aligned}\mathrm{D}_{\mu}^{\parallel}\vec{\Phi} & =\partial_{\mu}\left|\Phi\right|\\
 & =\mathrm{D}_{\mu}\left|\Phi\right|.
\end{aligned}
\end{equation}
\begin{figure}[H]
\noindent \begin{centering}
\includegraphics[scale=1.15]{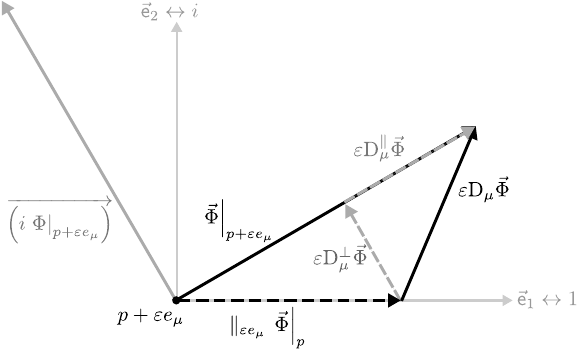}
\par\end{centering}
\caption{The internal space vector $\mathrm{D}_{\mu}\vec{\Phi}$ may be split
into components parallel and perpendicular to $\vec{\Phi}$. The length
of the parallel component is the orthogonal projection of $\mathrm{D}_{\mu}\vec{\Phi}$
on to $\vec{\Phi}$, which is the real inner product over the length
of $\vec{\Phi}$, i.e. $\mathrm{D}_{\mu}^{\parallel}\vec{\Phi}=\left\langle \mathrm{D}_{\mu}\vec{\Phi},\vec{\Phi}\right\rangle _{\mathbb{R}}/\left\Vert \vec{\Phi}\right\Vert $.
The length of the perpendicular component $\mathrm{D}_{\mu}^{\perp}\vec{\Phi}$
is the orthogonal projection of $\mathrm{D}_{\mu}\vec{\Phi}$ on to
the $\pi/2$-rotated $\vec{\Phi}$, which is $\left\langle \mathrm{D}_{\mu}\vec{\Phi},\protect\overrightarrow{\left(i\Phi\right)}\right\rangle _{\mathbb{R}}=\mathrm{Im}\left\langle \Phi,\mathrm{D}_{\mu}\Phi\right\rangle _{\mathbb{C}}$
over the length of $\vec{\Phi}$. In the figure we see that since
$\protect\overrightarrow{\left(i\Phi\right)}$ is rotated in the counterclockwise
direction, a positive value of $\mathrm{D}_{\mu}^{\perp}\vec{\Phi}$
represents a counterclockwise rotation of $\vec{\Phi}$ relative to
parallel transport.}
\end{figure}

Now, $\mathrm{D}_{\mu}^{\perp}\vec{\Phi}$ is the length of the component
perpendicular to $\vec{\Phi}$ of the difference between $\vec{\Phi}$
and its parallel transport in the $e_{\mu}$ direction, i.e. it is
the distance by which $\vec{\Phi}$ moves due to rotation, ignoring
the change in its length. Since the rotation is infinitesimal, the
sine is equal to the angle in radians; hence the angular velocity
of $\vec{\Phi}$ relative to parallel transport per unit distance
in the $e_{\mu}$ direction is 
\begin{align}
\begin{aligned}\mathrm{D}_{\mu}^{\varangle}\vec{\Phi} & \equiv\frac{\mathrm{D}_{\mu}^{\perp}\vec{\Phi}}{\left\Vert \vec{\Phi}\right\Vert }\\
 & =\frac{\mathrm{Im}\left\langle \Phi,\mathrm{D}_{\mu}\Phi\right\rangle _{\mathbb{C}}}{\left|\Phi\right|^{2}}\\
 & =\frac{\mathrm{Im}\left\langle \Phi,\partial_{\mu}\Phi\right\rangle _{\mathbb{C}}}{\left|\Phi\right|^{2}}-qA_{\mu}\\
 & =\frac{\partial_{\mu}^{\perp}\vec{\Phi}}{\left\Vert \vec{\Phi}\right\Vert }+A_{\mu}^{\varangle}\\
 & =\partial_{\mu}^{\varangle}\vec{\Phi}+A_{\mu}^{\varangle}
\end{aligned}
\end{align}
in radians per coordinate length. Note that as expected, this is the
counterclockwise angular velocity of the components in the chosen
gauge plus the counterclockwise angular velocity of the coordinate
axes relative to parallel transport. We can also see this by using
(\ref{eq:DPhi-components}), whereby the third line above yields
\begin{align}
\begin{aligned}\mathrm{D}_{\mu}^{\varangle}\Phi & =\frac{-i\left(\Phi^{*}\partial_{\mu}\Phi-\Phi\partial_{\mu}\Phi^{*}\right)}{2\left|\Phi\right|^{2}}-qA_{\mu}\\
 & =\partial_{\mu}\varphi-qA_{\mu}\\
 & =\mathrm{D}_{\mu}\varphi,
\end{aligned}
\end{align}
which again splits into the angular velocity of the coordinate plus
the angular velocity of the coordinate axes.

The above results allow us to use (\ref{eq:DPhiDPhi-complex-components})
to express the squared length of the perpendicular component in terms
of $\vec{\Phi}\in\mathbb{R}^{2}$ , i.e.
\begin{align}
\begin{aligned}\left\langle \mathrm{D}_{\mu}^{\perp}\vec{\Phi},\mathrm{D}_{\mu}^{\perp}\vec{\Phi}\right\rangle  & =\left\Vert \vec{\Phi}\right\Vert ^{2}\left\langle \mathrm{D}_{\mu}^{\varangle}\vec{\Phi},\mathrm{D}_{\mu}^{\varangle}\vec{\Phi}\right\rangle =\left|\Phi\right|^{2}\left\langle \mathrm{D}_{\mu}\varphi,\mathrm{D}_{\mu}\varphi\right\rangle \\
 & =\left\langle \mathrm{D}_{\mu}\Phi,\mathrm{D}_{\mu}\Phi\right\rangle -\left\langle \mathrm{D}_{\mu}\left|\Phi\right|,\mathrm{D}_{\mu}\left|\Phi\right|\right\rangle \\
 & =\left\langle \mathrm{D}_{\mu}\vec{\Phi},\mathrm{D}_{\mu}\vec{\Phi}\right\rangle -\left\langle \mathrm{D}_{\mu}^{\parallel}\vec{\Phi},\mathrm{D}_{\mu}^{\parallel}\vec{\Phi}\right\rangle \\
 & =\mathrm{D}_{\mu}\Phi^{a}\mathrm{D}^{\mu}\Phi_{a}-\frac{1}{\left\Vert \vec{\Phi}\right\Vert ^{2}}\left\langle \mathrm{D}_{\mu}\vec{\Phi},\vec{\Phi}\right\rangle _{\mathbb{R}}\left\langle \mathrm{D}^{\mu}\vec{\Phi},\vec{\Phi}\right\rangle _{\mathbb{R}}\\
 & =\mathrm{D}_{\mu}\Phi^{a}\mathrm{D}^{\mu}\Phi_{a}-\left(\Phi_{c}\Phi^{c}\right)^{-1}\Phi^{a}\left(\mathrm{D}_{\mu}\Phi_{a}\right)\Phi^{b}\left(\mathrm{D}^{\mu}\Phi_{b}\right).
\end{aligned}
\label{eq:DPerp2-real-expression}
\end{align}

Finally, multiplying by the area per radian yields
\begin{align}
\begin{aligned}\mathrm{D}_{\mu}^{\leftslice}\vec{\Phi} & \equiv\mathrm{D}_{\mu}^{\varangle}\vec{\Phi}\frac{\pi\left\Vert \vec{\Phi}\right\Vert ^{2}}{2\pi}\\
 & =\frac{1}{2}\left\Vert \vec{\Phi}\right\Vert ^{2}\mathrm{D}_{\mu}^{\varangle}\vec{\Phi}\\
 & =\frac{1}{2}\left\Vert \vec{\Phi}\right\Vert \mathrm{D}_{\mu}^{\perp}\vec{\Phi},
\end{aligned}
\end{align}
the (counterclockwise) area swept out by the matter field (relative
to parallel transport) per unit distance in the $e_{\mu}$ direction.
\begin{figure}[H]
\noindent \begin{centering}
\includegraphics[scale=1.15]{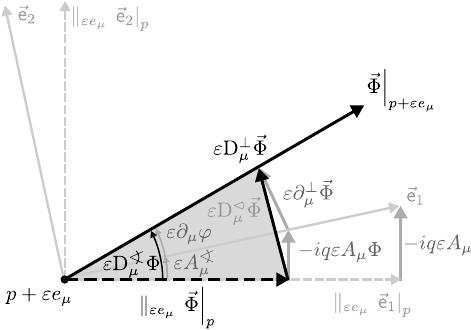}
\par\end{centering}
\caption{The infinitesimal counterclockwise rotation of the matter field vector
relative to parallel transport per unit distance in the $e_{\mu}$
direction is $\mathrm{D}_{\mu}^{\varangle}\Phi=\partial_{\mu}\varphi+A_{\mu}^{\varangle}$,
the rotation of the angle in $\Phi\equiv\left|\Phi\right|e^{i\varphi}$
plus the rotation of the coordinate axes. The infinitesimal area swept
out by the rotation of the matter field is $\mathrm{D}_{\mu}^{\leftslice}\vec{\Phi}=\frac{1}{2}\left\Vert \vec{\Phi}\right\Vert \mathrm{D}_{\mu}^{\perp}\vec{\Phi}$.}
\end{figure}

\subsection{Quasi-gauge electromagnetism}

Classical electromagnetism, excluding mass and gravity, is usually
defined (using natural units and the mostly pluses metric signature
in flat spacetime) as a $U(1)$ gauge theory with no matter field
and Lagrangian
\begin{align}
L_{\widetilde{\mathrm{EM}}} & \equiv A_{\mu}J_{q}^{\mu}-\frac{1}{4}F_{\mu\nu}F^{\mu\nu},
\end{align}
where 
\begin{equation}
J_{q}^{\mu}\equiv qJ^{\mu}\equiv q\rho_{0}U^{\mu}
\end{equation}
is the electromagnetic four-current, which may be defined in terms
of the charge $q$ and the matter four-current $J$, which in turn
may be defined in terms of the matter rest density $\rho_{0}$ and
the unit time-like four-vector $U$ in the direction of the four-current.
The charge $q$ may be viewed as being per particle number (theoretically
fractional for a continuum), which substitutes ``particle number''
for ``matter'' above. 

The equations of motion (EOM) which result from varying the gauge
potential are (see Section \ref{subsec:Gauge-potential-equations-of-motion},
or for a treatment in terms of forms see Section 4.1 in \cite{GockelerSchucker})
Maxwell's equations
\begin{align}
J_{q}^{\nu} & =\nabla_{\mu}F^{\nu\mu}\\
\Rightarrow\nabla_{\nu}J_{q}^{\nu} & =\nabla_{\nu}\nabla_{\mu}F^{\nu\mu}=0,\label{eq:divJ=00003D0}
\end{align}
where $\nabla$ is the Levi-Civita covariant derivative, so that (\ref{eq:divJ=00003D0})
implies $J_{q}$ is divergenceless. These EOM are used to excuse the
fact that the Lagrangian is not gauge invariant (and is therefore
ill-defined from the geometric point of view), since under a gauge
transformation $e^{-iq\Lambda}$ we have 
\begin{equation}
\begin{aligned}A_{\mu} & \rightarrow A_{\mu}+\partial_{\mu}\Lambda\\
\Rightarrow L_{\widetilde{\mathrm{EM}}} & \rightarrow L_{\widetilde{\mathrm{EM}}}-J_{q}^{\mu}\partial_{\mu}\Lambda\\
 & =L_{\widetilde{\mathrm{EM}}}-\nabla_{\mu}\left(J_{q}^{\mu}\Lambda\right)+\Lambda\nabla_{\mu}J_{q}^{\mu},
\end{aligned}
\end{equation}
where the first extra term is a divergence and therefore does not
change the equations of motion, while the second vanishes via the
EOM. Varying the metric (see Section \ref{subsec:Metric-equations-of-motion})
identifies the Hilbert stress energy momentum (SEM) tensor as
\begin{align}
T_{\mathrm{EM}}^{\mu\nu} & \equiv F^{\mu\lambda}F^{\nu}{}_{\lambda}-\frac{1}{4}F^{\lambda\sigma}F_{\lambda\sigma}g^{\mu\nu}.\label{eq:em-field-hilbert-sem-tensor}
\end{align}

Following e.g. \cite{Dirac}, we may include gravity and associate
a mass $m$ as well as a charge with the four-current by defining
\begin{align}
L_{\mathrm{G}\widetilde{\mathrm{EM}}} & \equiv-m\sqrt{-J^{\mu}J^{\nu}g_{\mu\nu}}+qA_{\mu}J^{\mu}-\frac{1}{4}g^{\mu\nu}g^{\lambda\sigma}F_{\mu\lambda}F_{\nu\sigma}+\frac{1}{16\pi}R,
\end{align}
where $R$ is the spacetime scalar curvature. The new terms leave
the gauge potential EOM unchanged, but upon variation of the metric
yield a Hilbert SEM tensor with a time-like dust term 
\begin{align}
T_{\mathrm{GEM}}^{\mu\nu} & \equiv m\rho_{0}U^{\mu}U^{\nu}+T_{\mathrm{EM}}^{\mu\nu}.\label{eq:gravity-quasi-em-SEM-tensor}
\end{align}
This tensor is proportional to the Einstein tensor and hence must
be divergenceless, which (see Section \ref{subsec:Metric-equations-of-motion})
yields the equations of geodesic deviation
\begin{equation}
m\rho_{0}\left(\nabla_{U}U\right)^{\mu}=F^{\mu}{}_{\nu}J_{q}^{\nu},
\end{equation}
better known as the Lorentz force law. 

\subsection{Klein-Gordon theory}

For reference, we here summarize Klein-Gordon theory (complex, minimally
coupled, and in flat spacetime; sometimes then called scalar electrodynamics
or scalar QED), which is usually defined (again using natural units
and the mostly pluses metric signature) as a $U(1)$ gauge theory
with a complex scalar matter field and Lagrangian
\begin{align}
L_{\mathrm{KG}} & \equiv-\left(\mathrm{D}_{\mu}\Phi\right)^{*}\mathrm{D}^{\mu}\Phi-m^{2}\Phi^{*}\Phi-\frac{1}{4}F_{\mu\nu}F^{\mu\nu}.
\end{align}
This is of course usually considered as a quantum theory, but here
we want to explore whether it can provide at least inspiration for
a model of classical electromagnetism. 

Expanding the first and last terms (see Section \ref{subsec:The-Lagrangian}),
we find the EOM from varying the gauge potential are 
\begin{equation}
\begin{aligned}-iq\left(\Phi^{*}\mathrm{D}^{\nu}\Phi-\Phi\left(\mathrm{D}^{\nu}\Phi\right)^{*}\right) & =\nabla_{\mu}F^{\nu\mu},\end{aligned}
\label{eq:kg-gp-eom}
\end{equation}
allowing us to identify the divergenceless matter four-current
\begin{equation}
\begin{aligned}J^{\nu} & \equiv-i\left(\Phi^{*}\mathrm{D}^{\nu}\Phi-\Phi\left(\mathrm{D}^{\nu}\Phi\right)^{*}\right),\end{aligned}
\end{equation}
in terms of which (\ref{eq:kg-gp-eom}) is Maxwell's equations. Varying
the matter field, however, results in the Klein-Gordon equation 
\begin{align}
\mathrm{D}_{\mu}\mathrm{D}^{\mu}\Phi & =m^{2}\Phi,
\end{align}
from which point it is difficult to proceed classically. Firstly,
these EOM do not constrain the particle density associated with $J$
to be positive, and secondly, we have no way to eliminate $\Phi$
from the Lagrangian in order to obtain the classical Hilbert SEM tensor
$T_{\mathrm{GEM}}^{\mu\nu}$ of (\ref{eq:gravity-quasi-em-SEM-tensor}).

We nevertheless may obtain some results which might act as inspiration
for a classical theory by taking $A=0$. Then (free complex) Klein-Gordon
theory no longer includes electromagnetism, but the associated Klein-Gordon
equation
\begin{align}
\begin{aligned}\partial_{\mu}\partial^{\mu}\Phi & =m^{2}\Phi\end{aligned}
\end{align}
has a general solution which is a linear combination of plane wave
solutions
\begin{align}
\Phi_{P} & \equiv\left|\Phi\right|e^{iP_{\mu}x^{\mu}},
\end{align}
where the complex modulus $\left|\Phi\right|$ and the four-vector
$P$ are constant. Taking derivatives, we have
\begin{align}
\begin{aligned}\partial_{\mu}\Phi_{P} & =iP_{\mu}\Phi_{P}\\
\Rightarrow\partial_{\mu}\partial^{\mu}\Phi_{P} & =-P_{\mu}P^{\mu}\Phi_{P}=m^{2}\Phi_{P}\\
\Rightarrow P^{\mu} & =mU^{\mu},
\end{aligned}
\end{align}
where $U$ is again a unit time-like four-vector, while
\begin{align}
\begin{aligned}J_{P}^{\mu} & =2\left|\Phi\right|^{2}P^{\mu}\\
\Rightarrow\rho_{0} & =2m\left|\Phi\right|^{2}.
\end{aligned}
\end{align}
Geometrically, these are matter fields which rotate in the time-like
$P^{\mu}$ direction with an angular velocity of $m$ radians per
unit proper time, and whose constant length squared is proportional
to the rest density. Moreover, on-shell for a plane wave the Lagrangian
may be written as
\begin{align}
L_{\mathrm{KG}-P} & =-\frac{1}{2}m\rho_{0}U_{\mu}U_{\nu}g^{\mu\nu}-\frac{1}{2}m\rho_{0},
\end{align}
which upon varying the metric (see Section \ref{subsec:Metric-equations-of-motion})
yields a time-like dust Hilbert SEM tensor of $m\rho_{0}U^{\mu}U^{\nu}$. 

This is all quite suggestive, but faces two issues as inspiration
for constructing a classical gauge theory of electromagnetism: (1)
we eliminated the electromagnetic interaction to get solutions, and
(2) we cannot construct a general composite classical four-current
by combining plane wave solutions, since their phases interfere with
each other, i.e. they are quantum in nature. In Section \ref{sec:Matter-field-electromagnetism}
which follows, we arrive at an alternative Lagrangian which avoids
these issues while taking advantage of the above observations.

\section{\label{sec:Matter-field-electromagnetism}Matter field electromagnetism }

\subsection{\label{subsec:The-Lagrangian}The Lagrangian}

With the geometric quantities of Section \ref{sec:Geometric-U1-gauge-theory}
in hand, we now define matter field electromagnetism (MFEM) (again
using natural units and the mostly pluses metric signature) as a geometric
$U(1)$ gauge theory with Lagrangian
\begin{align}
\begin{aligned}L_{\mathrm{EM}} & \equiv-\left\langle \mathrm{D}_{\mu}^{\perp}\vec{\Phi},\mathrm{D}_{\mu}^{\perp}\vec{\Phi}\right\rangle -m^{2}\left\langle \vec{\Phi},\vec{\Phi}\right\rangle -\frac{1}{2}\left\langle F,F\right\rangle \\
 & =\frac{1}{4\left|\Phi\right|^{2}}\left(\Phi^{*}\mathrm{D}_{\mu}\Phi-\Phi\left(\mathrm{D}_{\mu}\Phi\right)^{*}\right)\left(\Phi^{*}\mathrm{D}^{\mu}\Phi-\Phi\left(\mathrm{D}^{\mu}\Phi\right)^{*}\right)-m^{2}\left|\Phi\right|^{2}-\frac{1}{4}F_{\mu\nu}F^{\mu\nu}.
\end{aligned}
\label{eq:MFEM-Lagrangian}
\end{align}
Geometrically, the absolute value of the dynamical term is the squared
perpendicular distance between the matter field and its parallel transport
per unit distance in the direction in which this distance is extremal.
Its negative can be written in terms of various quantities as
\begin{align}
\begin{aligned}-L_{\mathrm{EM-D}} & =\left\Vert \mathrm{D}_{\mu}^{\perp}\vec{\Phi}\right\Vert ^{2}\\
 & =\left\Vert \vec{\Phi}\right\Vert ^{2}\left\langle \mathrm{D}_{\mu}^{\varangle}\vec{\Phi},\mathrm{D}_{\mu}^{\varangle}\vec{\Phi}\right\rangle \\
 & =\left\Vert \mathrm{D}_{\mu}\vec{\Phi}\right\Vert ^{2}-\frac{1}{\left\Vert \vec{\Phi}\right\Vert ^{2}}\left\langle \mathrm{D}_{\mu}\vec{\Phi},\vec{\Phi}\right\rangle _{\mathbb{R}}\left\langle \mathrm{D}^{\mu}\vec{\Phi},\vec{\Phi}\right\rangle _{\mathbb{R}}\\
 & =\mathrm{D}_{\mu}\Phi^{a}\mathrm{D}^{\mu}\Phi_{a}-\left(\Phi_{c}\Phi^{c}\right)^{-1}\Phi^{a}\left(\mathrm{D}_{\mu}\Phi_{a}\right)\Phi^{b}\left(\mathrm{D}^{\mu}\Phi_{b}\right)\\
 & =\frac{1}{\left|\Phi\right|^{2}}\mathrm{Im}\left\langle \mathrm{D}_{\mu}\Phi,\Phi\right\rangle _{\mathbb{C}}\mathrm{Im}\left\langle \mathrm{D}^{\mu}\Phi,\Phi\right\rangle _{\mathbb{C}}\\
 & =\left|\Phi\right|^{2}\left(\partial_{\mu}\varphi-qA_{\mu}\varphi\right)\left(\partial^{\mu}\varphi-qA^{\mu}\varphi\right).
\end{aligned}
\label{eq:dynamical-term-expressions}
\end{align}
Note that if $\mathrm{D}_{\mu}^{\parallel}\vec{\Phi}=0$, i.e. if
the matter field has constant length, then $\mathrm{D}_{\mu}^{\perp}\vec{\Phi}=\mathrm{D}_{\mu}\vec{\Phi}$
and our Lagrangian is identical to the Klein-Gordon Lagrangian.

In order to vary the gauge potential, we must express the Lagrangian
explicitly in terms of $A_{\mu}$. Expanding out the gauge covariant
derivatives yields a dynamical term for gauge potential variations
of
\begin{align}
\begin{aligned}L_{\mathrm{EM-D}} & =\frac{1}{4\left|\Phi\right|^{2}}\left\Vert \Phi^{*}\partial_{\mu}\Phi-\Phi\partial_{\mu}\Phi^{*}-2iq\left|\Phi\right|^{2}A_{\mu}\right\Vert ^{2}\\
\Rightarrow L_{\mathrm{EM-D}}\left(A_{\mu}\right) & =-iqA_{\mu}\left(\Phi^{*}\partial^{\mu}\Phi-\Phi\partial^{\mu}\Phi^{*}\right)-q^{2}\left|\Phi\right|^{2}A_{\mu}A^{\mu}.
\end{aligned}
\label{eq:L_EM-D(A)}
\end{align}
Note that this is identical to that of the Klein-Gordon dynamical
term
\begin{align}
\begin{aligned}L_{\mathrm{KG-D}} & =-\left(\partial_{\mu}\Phi-iqA_{\mu}\Phi\right)\left(\partial^{\mu}\Phi^{*}+iqA^{\mu}\Phi^{*}\right)\\
\Rightarrow L_{\mathrm{KG-D}}\left(A_{\mu}\right) & =-iqA_{\mu}\left(\Phi^{*}\partial^{\mu}\Phi-\Phi\partial^{\mu}\Phi^{*}\right)-q^{2}\left|\Phi\right|^{2}A_{\mu}A^{\mu},
\end{aligned}
\end{align}
which means we will end up with the same gauge potential EOM.

\subsection{\label{subsec:Matter-field-equations-of-motion}Matter field equations
of motion}

The Euler-Lagrange equation may be written 
\begin{align}
\frac{\partial L}{\partial\Phi} & =\mathrm{D}_{\mu}p_{\Phi}^{\mu}
\end{align}
in terms of the canonical momentum
\begin{align}
p_{\Phi}^{\mu} & \equiv\frac{\partial L}{\partial\left(\mathrm{D}_{\mu}\Phi\right)}.
\end{align}
In keeping with our geometrical viewpoint, we opt to work here in
terms of the real vector components $\Phi^{a}$, but in Section \ref{subsec:Noether's-theorem}
we derive the same results in terms of $\Phi$ and $\Phi^{*}$ as
independent quantities, as is more common. Recalling (\ref{eq:DPerp2-real-expression}),
we have
\begin{align}
\begin{aligned}L_{\mathrm{EM}}\left(\Phi^{a},\mathrm{D}_{\mu}\Phi^{a}\right) & =-\mathrm{D}_{\mu}\Phi^{a}\mathrm{D}^{\mu}\Phi_{a}+\left(\Phi_{c}\Phi^{c}\right)^{-1}\Phi^{a}\left(\mathrm{D}_{\mu}\Phi_{a}\right)\Phi^{b}\left(\mathrm{D}^{\mu}\Phi_{b}\right)-m^{2}\Phi_{a}\Phi^{a},\\
\frac{\partial L_{\mathrm{EM}}}{\partial\Phi^{d}} & =-2\left(\Phi_{c}\Phi^{c}\right)^{-2}\Phi_{d}\Phi^{a}\left(\mathrm{D}_{\mu}\Phi_{a}\right)\Phi^{b}\left(\mathrm{D}^{\mu}\Phi_{b}\right)\\
 & \phantom{{}=}+2\left(\Phi_{c}\Phi^{c}\right)^{-1}\Phi^{a}\left(\mathrm{D}_{\mu}\Phi_{a}\right)\left(\mathrm{D}^{\mu}\Phi_{d}\right)-2m^{2}\Phi_{d},\\
p_{\Phi^{d}}^{\mu}=\frac{\partial L_{\mathrm{EM}}}{\partial\left(\mathrm{D}_{\mu}\Phi^{d}\right)} & =-2\mathrm{D}^{\mu}\Phi_{d}+2\left(\Phi_{c}\Phi^{c}\right)^{-1}\Phi^{a}\left(\mathrm{D}^{\mu}\Phi_{a}\right)\Phi_{d}\\
\Rightarrow\mathrm{D}_{\mu}p_{\Phi^{d}}^{\mu} & =-2\mathrm{D}_{\mu}\mathrm{D}^{\mu}\Phi_{d}-4\left(\Phi_{c}\Phi^{c}\right)^{-2}\Phi^{b}\left(\mathrm{D}_{\mu}\Phi_{b}\right)\Phi^{a}\left(\mathrm{D}^{\mu}\Phi_{a}\right)\Phi_{d}\\
 & \phantom{{}=}+2\left(\Phi_{c}\Phi^{c}\right)^{-1}\mathrm{D}_{\mu}\Phi^{a}\left(\mathrm{D}^{\mu}\Phi_{a}\right)\Phi_{d}\\
 & \phantom{{}=}+2\left(\Phi_{c}\Phi^{c}\right)^{-1}\Phi^{a}\left(\mathrm{D}_{\mu}\mathrm{D}^{\mu}\Phi_{a}\right)\Phi_{d}+2\left(\Phi_{c}\Phi^{c}\right)^{-1}\Phi^{a}\left(\mathrm{D}^{\mu}\Phi_{a}\right)\mathrm{D}_{\mu}\Phi_{d}.
\end{aligned}
\label{eq:matter-field-derivatives}
\end{align}
But 
\begin{equation}
2\left(\Phi_{c}\Phi^{c}\right)^{-1}\Phi^{a}\left(\mathrm{D}_{\mu}\mathrm{D}^{\mu}\Phi_{a}\right)\Phi_{d}=2\left(\mathrm{D}_{\mu}\mathrm{D}^{\mu}\Phi_{d}\right)
\end{equation}
since applying each side to $\Phi^{d}$ yields the same result. Therefore
the Euler-Lagrange equation is
\begin{align}
\begin{aligned}-2m^{2}\Phi_{d} & =-2\left(\Phi_{c}\Phi^{c}\right)^{-2}\Phi^{b}\left(\mathrm{D}_{\mu}\Phi_{b}\right)\Phi^{a}\left(\mathrm{D}^{\mu}\Phi_{a}\right)\Phi_{d}+2\left(\Phi_{c}\Phi^{c}\right)^{-1}\mathrm{D}_{\mu}\Phi^{a}\mathrm{D}^{\mu}\Phi_{a}\Phi_{d}\\
 & =-2\left(\Phi_{c}\Phi^{c}\right)^{-1}\Phi_{d}\left(-\mathrm{D}_{\mu}\Phi^{a}\mathrm{D}^{\mu}\Phi_{a}+\left(\Phi_{c}\Phi^{c}\right)^{-1}\Phi^{a}\left(\mathrm{D}^{\mu}\Phi_{a}\right)\Phi^{b}\left(\mathrm{D}_{\mu}\Phi_{b}\right)\right)\\
\Rightarrow-m^{2} & =\left\langle \mathrm{D}_{\mu}^{\varangle}\vec{\Phi},\mathrm{D}_{\mu}^{\varangle}\vec{\Phi}\right\rangle .
\end{aligned}
\end{align}
We obtain this same result if, as is more common, we use the equivalent
Euler-Lagrange expression $\partial L/\partial\Phi=\partial_{\mu}\left(\partial L/\partial\left(\partial_{\mu}\Phi\right)\right)$
in flat spacetime. 

Using the results of Section \ref{subsec:Inner-products}, we may
geometrically describe the solutions of this equation as those for
which at every point in spacetime, the unit length four-vector $U^{\mu}$
in the direction of extremal angular velocity is time-like, and in
this direction the angular velocity is constant and equal to $m$:
\begin{equation}
\mathrm{D}_{U}^{\varangle}\vec{\Phi}=\pm m
\end{equation}
So when the matter field satisfies the EOM (is on-shell), $m$ is
identified with the (constant) angular velocity of the matter field
(relative to parallel transport) per unit proper time in the direction
for which this angular velocity is extremal. In Section \ref{subsec:Metric-equations-of-motion},
$m$ will be also identified as the mass associated with the matter
four-current, which in Section \ref{subsec:Gauge-potential-equations-of-motion}
we will see points in this same direction.

We may also note that the matter field part of $L_{\mathrm{EM}}$
vanishes on-shell (as expected from Euler's homogeneous function theorem,
since it is homogeneous of degree 2 and $\nabla_{\mu}\left(\Phi^{d}p_{\Phi^{d}}^{\mu}\right)=0$);
and we now have yet another way to express the negative dynamical
term, this time as the square of the canonical momentum
\begin{align}
\begin{aligned}-4L_{\mathrm{EM-D}} & =\left\Vert p_{\Phi}\right\Vert ^{2},\end{aligned}
\end{align}
reminiscent of Lagrangians for relativistic particles (the canonical
momentum obtained in Section \ref{subsec:Noether's-theorem} by differentiating
by $\Phi$ independently of $\Phi^{*}$ even removes the factor of
four).

\subsection{\label{subsec:Gauge-potential-equations-of-motion}Gauge potential
equations of motion}

From (\ref{eq:L_EM-D(A)}) and (\ref{eq:FF}), the Lagrangian for
gauge potential variations is 
\begin{align}
\begin{aligned}L_{\mathrm{EM}}\left(A_{\mu},\partial_{\nu}A_{\mu}\right) & =-iqA_{\mu}\left(\Phi^{*}\partial^{\mu}\Phi-\Phi\partial^{\mu}\Phi^{*}\right)-q^{2}\left|\Phi\right|^{2}A_{\mu}A^{\mu}\\
 & \phantom{{}=}-\frac{1}{2}\left(\partial_{\mu}A_{\nu}\partial^{\mu}A^{\nu}-\partial_{\mu}A_{\nu}\partial^{\nu}A^{\mu}\right).
\end{aligned}
\end{align}
Our dynamical term does not depend on the derivative of the gauge
potential, so the electromagnetic four-current will be $\partial L/\partial A_{\nu}$
as is usual in other EM gauge theories. Explicitly,
\begin{equation}
\begin{aligned}\frac{\partial L}{\partial A_{\nu}} & =-iq\left(\Phi^{*}\partial^{\nu}\Phi-\Phi\partial^{\nu}\Phi^{*}\right)-2q^{2}\left|\Phi\right|^{2}A^{\nu}\\
 & =-iq\left(\Phi^{*}\mathrm{D}^{\nu}\Phi-\Phi\left(\mathrm{D}^{\nu}\Phi\right)^{*}\right),\\
p_{A_{\nu}}^{\mu}\equiv\frac{\partial L}{\partial\left(\partial_{\mu}A_{\nu}\right)} & =-\left(\partial^{\mu}A^{\nu}-\partial^{\nu}A^{\mu}\right)\\
 & =-F^{\mu\nu},
\end{aligned}
\label{eq:A-derivatives}
\end{equation}
yielding an Euler-Lagrange equation of
\begin{equation}
\begin{aligned}iq\left(\Phi^{*}\mathrm{D}^{\nu}\Phi-\Phi\left(\mathrm{D}^{\nu}\Phi\right)^{*}\right) & =\nabla_{\mu}F^{\mu\nu},\end{aligned}
\end{equation}
which identifies the electromagnetic four-current as
\begin{equation}
\begin{aligned}J_{q}^{\nu} & \equiv-iq\left(\Phi^{*}\mathrm{D}^{\nu}\Phi-\Phi\left(\mathrm{D}^{\nu}\Phi\right)^{*}\right).\end{aligned}
\end{equation}
As in (\ref{eq:divJ=00003D0}) $J_{q}$ is therefore divergenceless
when the gauge potential EOM are satisfied; reversing the indices
of $F$ yields the usual form of Maxwell's equations
\begin{equation}
\begin{aligned}J_{q}^{\nu} & =\nabla_{\mu}F^{\nu\mu}.\end{aligned}
\end{equation}

\subsection{\label{subsec:Geometry-of-the-four-current}Geometry of the four-current}

The matter four-current may be written in terms of the various quantities
we have defined as
\begin{equation}
\begin{aligned}J_{\mu} & =-i\left(\Phi^{*}\mathrm{D}_{\nu}\Phi-\Phi\left(\mathrm{D}_{\nu}\Phi\right)^{*}\right)\\
 & =-i\left(\Phi^{*}\partial_{\mu}\Phi-\Phi\partial_{\mu}\Phi^{*}-2iq\left|\Phi\right|^{2}A_{\mu}\right)\\
 & =2\left\Vert \vec{\Phi}\right\Vert \mathrm{D}_{\mu}^{\perp}\vec{\Phi}\\
 & =2\left\Vert \vec{\Phi}\right\Vert ^{2}\mathrm{D}_{\mu}^{\varangle}\vec{\Phi}\\
 & =4\mathrm{D}_{\mu}^{\leftslice}\vec{\Phi}\\
 & =2\left|\Phi\right|^{2}\left(\partial_{\mu}\varphi+A_{\mu}^{\varangle}\right)\\
 & =2\mathrm{Im}\left\langle \mathrm{D}_{\mu}\Phi,\Phi\right\rangle _{\mathbb{C}}.
\end{aligned}
\label{eq:four-current-definitions}
\end{equation}
The fourth expression in (\ref{eq:four-current-definitions}) explicitly
identifies the direction of the four-current at a given point as being
the direction in which the matter field angular velocity (relative
to parallel transport) is extremal. Recall from Section \ref{subsec:Matter-field-equations-of-motion}
that the matter field EOM require the unit length four-vector $U^{\mu}$
in this direction to be time-like. We may then write 
\begin{equation}
\begin{aligned}J_{\mu} & =\rho_{0}U_{\mu}\\
\Rightarrow\rho_{0} & =-U^{\mu}J_{\mu}\\
 & =-2\left\Vert \vec{\Phi}\right\Vert ^{2}\mathrm{D}_{U}^{\varangle}\vec{\Phi}\\
 & =-4\mathrm{D}_{U}^{\leftslice}\vec{\Phi},
\end{aligned}
\label{eq:off-shell-current}
\end{equation}
so that geometrically the matter per volume orthogonal to the four-current
is proportional to the internal space area swept out clockwise by
the matter field per unit proper time along the four-current. 

Since the matter field EOM are $\mathrm{D}_{U}^{\varangle}\vec{\Phi}=\pm m$,
and particle density must be positive, we henceforth only consider
the negative value 
\begin{equation}
\mathrm{D}_{U}^{\varangle}\vec{\Phi}=-m
\end{equation}
to be physical, so that we have
\begin{equation}
\begin{aligned}\begin{aligned}\rho_{0} & =2\left\Vert \vec{\Phi}_{\mathrm{EL}}\right\Vert ^{2}m\\
\Rightarrow J^{\mu} & =2\left\Vert \vec{\Phi}_{\mathrm{EL}}\right\Vert ^{2}mU^{\mu}
\end{aligned}
\end{aligned}
\label{eq:on-shell-current-via-matter-field}
\end{equation}
in terms of the on-shell matter field $\vec{\Phi}_{\mathrm{EL}}$,
whose rotation relative to parallel transport in the direction $U$
is always clockwise. The reversed sign matter field EOM may be associated
with a positive particle density by defining the electromagnetic four-current
with the opposite sign, which is equivalent to simply flipping the
sign of $q$. Thus positively charged four-currents rotate clockwise,
while negatively charged four-currents rotate counterclockwise. 

We presently see an immediate distinction with the classical theory
of continua: the metric dependence of $\rho_{0}$ and $J^{\mu}$.
In the classical theory, $\rho_{0}$ and $J^{\mu}$ both depend upon
the metric such that the four-current density $\mathfrak{J}\equiv J\sqrt{-\det\left(g_{\mu\nu}\right)}$
is metric-independent. In MFEM, $\rho_{0}$ has a different metric
dependency and $J$ is a metric-independent 1-form, except on-shell
where $\rho_{0}$ is metric-independent and $J$ is proportional to
a unit vector. This will be explained in Section \ref{subsec:The-discrete-current}
when viewing MFEM as an approximation of QED. 

It is important to note that despite the appearance of $m$ in the
on-shell expression for $J$, it is the expression for the matter
(particle number) four-current; the electromagnetic four-current is
$J_{q}\equiv qJ$, just as the mass four-current is $J_{m}\equiv mJ$,
which will therefore have a factor $m^{2}$. We also may note that
we can now write the negative dynamical term of the Lagrangian on-shell
as
\begin{align}
\begin{aligned}-L_{\mathrm{EM-D-EL}} & =\frac{1}{4\left\Vert \vec{\Phi}_{\mathrm{EL}}\right\Vert ^{2}}\left\langle J,J\right\rangle \\
 & =\frac{1}{2}\frac{m}{\rho_{0}}\left\langle J,J\right\rangle \\
 & =\frac{1}{2}m\rho_{0}\left\langle U,U\right\rangle ,
\end{aligned}
\label{eq:dynamic-term-via-current}
\end{align}
which is the form of a Lagrangian for relativistic dust.

\subsection{\label{subsec:Metric-equations-of-motion}Metric equations of motion}

As noted in the previous section, both the four-current 1-form $J_{\nu}=-i\left(\Phi^{*}\mathrm{D}_{\nu}\Phi-\Phi\left(\mathrm{D}_{\nu}\Phi\right)^{*}\right)$
and the rest density $\rho_{0}=2\left\Vert \vec{\Phi}_{\mathrm{EL}}\right\Vert ^{2}m$
are metric-independent, so that for an on-shell matter field we use
(\ref{eq:dynamic-term-via-current}) and (\ref{eq:on-shell-current-via-matter-field})
to write the action as
\begin{equation}
\begin{aligned}S_{\mathrm{EM}}\left(g_{\mu\nu}\right) & =\int\left(-\frac{1}{2}\frac{m}{\rho_{0}}J_{\mu}J_{\nu}g^{\mu\nu}-\frac{1}{2}m\rho_{0}-\frac{1}{4}g^{\mu\nu}g^{\lambda\sigma}F_{\mu\lambda}F_{\nu\sigma}\right)\sqrt{g}\mathrm{d}^{4}x.\end{aligned}
\end{equation}
Recalling that $\delta g^{\mu\nu}=-g^{\mu\lambda}g^{\nu\sigma}\delta g_{\lambda\sigma}$
and $\delta\left(\sqrt{g}\right)=\frac{1}{2}\sqrt{g}g^{\mu\nu}\delta g_{\mu\nu}$,
the variation of the action yields
\begin{equation}
\begin{aligned}\delta S_{\mathrm{EM}}\left(g_{\mu\nu}\right) & =\frac{1}{2}\int\left(\frac{m}{\rho_{0}}J^{\mu}J^{\nu}\sqrt{g}-\frac{m}{\rho_{0}}J_{\lambda}J_{\sigma}g^{\lambda\sigma}\frac{1}{2}\sqrt{g}g^{\mu\nu}-m\rho_{0}\frac{1}{2}\sqrt{g}g^{\mu\nu}\right)\delta g_{\mu\nu}\mathrm{d}^{4}x\\
 & \phantom{{}=}-\frac{1}{4}\int\left(2g^{\lambda\sigma}F_{\mu\lambda}F_{\nu\sigma}\sqrt{g}\delta g^{\mu\nu}+F^{\lambda\sigma}F_{\lambda\sigma}\frac{1}{2}\sqrt{g}g^{\mu\nu}\delta g_{\mu\nu}\right)\mathrm{d}^{4}x\\
 & =\frac{1}{2}\int\left(m\rho_{0}U^{\mu}U^{\nu}+T_{\mathrm{EM}}^{\mu\nu}\right)\delta g_{\mu\nu}\sqrt{g}\mathrm{d}^{4}x,
\end{aligned}
\end{equation}
where $T_{\mathrm{EM}}^{\mu\nu}$ is defined per (\ref{eq:em-field-hilbert-sem-tensor})
and the quantity in parentheses in the last line defines the Hilbert
SEM tensor, which is equal to $T_{\mathrm{GEM}}^{\mu\nu}$ from (\ref{eq:gravity-quasi-em-SEM-tensor}),
matching that of quasi-gauge EM as desired. 

This implies the Lorentz force law, completing our equivalence with
classical electromagnetism, as we quickly review (see e.g. \cite{Dirac}).
$T_{\mathrm{GEM}}^{\mu\nu}$ is proportional to the Einstein tensor
and hence must be divergenceless, which yields the equations of geodesic
deviation
\begin{equation}
\begin{aligned}\nabla_{\nu}T_{\mathrm{GEM}}^{\mu\nu} & =\nabla_{\nu}\left(m\rho_{0}U^{\mu}U^{\nu}+g_{\lambda\sigma}F^{\mu\lambda}F^{\nu\sigma}-\frac{1}{4}F^{\lambda\sigma}F_{\lambda\sigma}g^{\mu\nu}\right)\\
 & =mU^{\mu}\nabla_{\nu}J^{\nu}+m\rho_{0}U^{\nu}\nabla_{\nu}U^{\mu}\\
 & \phantom{{}=}+g_{\lambda\sigma}F^{\mu\lambda}\nabla_{\nu}F^{\nu\sigma}+g_{\lambda\sigma}F^{\nu\sigma}\nabla_{\nu}F^{\mu\lambda}-\frac{1}{2}F^{\lambda\sigma}\nabla_{\nu}F_{\lambda\sigma}g^{\mu\nu}\\
 & =m\rho_{0}\left(\nabla_{U}U\right)^{\mu}+F^{\mu}{}_{\sigma}\nabla_{\nu}F^{\nu\sigma}\\
 & \phantom{{}=}+\frac{1}{2}F^{\lambda\sigma}g^{\mu\nu}\left(\nabla_{\lambda}F_{\nu\sigma}-\nabla_{\sigma}F_{\nu\lambda}-\nabla_{\nu}F_{\lambda\sigma}\right)\\
\Rightarrow m\rho_{0}\left(\nabla_{U}U\right)^{\mu} & =F^{\mu}{}_{\sigma}J_{q}^{\sigma},
\end{aligned}
\end{equation}
where in the penultimate equality we use $\nabla_{\nu}J^{\nu}=0$
and the anti-symmetry of $F$, yielding the three terms in parentheses
which vanish due to the second Bianchi identity, and in the last line
we use the gauge potential EOM. $U$ is the unit four-vector in the
direction of the proper time $\tau$ of the four-current, so that
at a point in flat spacetime this equation becomes
\begin{equation}
\begin{aligned}\partial_{\tau}P^{\mu} & =qF^{\mu}{}_{\sigma}U^{\sigma},\end{aligned}
\end{equation}
where $P=mU$. In an inertial frame we then have components 
\begin{equation}
\begin{aligned}\partial_{\tau}\begin{pmatrix}E & p^{x} & p^{y} & p^{z}\end{pmatrix} & =q\begin{pmatrix}0 & E^{x} & E^{y} & E^{z}\\
E^{x} & 0 & B^{z} & -B^{y}\\
E^{y} & -B^{z} & 0 & B^{x}\\
E^{z} & B^{y} & -B^{x} & 0
\end{pmatrix}\begin{pmatrix}\gamma\\
\gamma v^{x}\\
\gamma v^{y}\\
\gamma v^{z}
\end{pmatrix}\\
\Rightarrow\partial_{t}\left(\mathbf{p}\right) & =q\left(\mathbf{E}+\mathbf{v}\times\mathbf{B}\right),
\end{aligned}
\end{equation}
where $\gamma$ is the Lorentz factor, we use $\partial_{\tau}=\gamma\partial_{t}$,
and recall that $\mathbf{p}\equiv\gamma m\mathbf{v}$ is the relativistic
momentum, which in the non-relativistic limit is just the momentum. 

\subsection{\label{subsec:Noether's-theorem}Noether's theorem}

In order to take advantage of the simplicity of complex gauge transformations,
we here re-derive the matter field EOM in complex notation. As with
similar theories, we may vary $\Phi$ and $\Phi^{*}$ as independent
quantities. Using the complex second expression of (\ref{eq:MFEM-Lagrangian}),
we write the Lagrangian for matter field variations in terms of the
four-current as 
\begin{align}
\begin{aligned}L_{\mathrm{EM}}\left(\Phi,\mathrm{D}_{\mu}\Phi\right) & =\frac{1}{4\Phi^{*}\Phi}\left(iJ_{\mu}\right)\left(iJ^{\mu}\right)-m^{2}\Phi^{*}\Phi,\\
iJ_{\mu} & =\Phi^{*}\mathrm{D}_{\mu}\Phi-\Phi\left(\mathrm{D}_{\mu}\Phi\right)^{*}.
\end{aligned}
\end{align}
Calculating derivatives then yields
\begin{align}
\begin{aligned}\frac{\partial L_{\mathrm{EM}}}{\partial\Phi} & =\frac{J_{\mu}J^{\mu}\Phi^{*}}{4\left(\Phi^{*}\Phi\right)^{2}}-\frac{iJ^{\mu}\left(\mathrm{D}_{\mu}\Phi\right)^{*}}{2\Phi^{*}\Phi}-m^{2}\Phi^{*},\\
p_{\Phi}^{\mu}=\frac{\partial L_{\mathrm{EM}}}{\partial\left(\mathrm{D}_{\mu}\Phi\right)} & =\frac{\Phi^{*}iJ^{\mu}}{2\Phi^{*}\Phi}\\
\Rightarrow\mathrm{D}_{\mu}\left(p_{\Phi}^{\mu}\right) & =-\frac{\Phi^{*}iJ^{\mu}}{2\left(\Phi^{*}\Phi\right)^{2}}\partial_{\mu}\left|\Phi\right|^{2}+\frac{\left(\mathrm{D}_{\mu}\Phi\right)^{*}iJ^{\mu}}{2\Phi^{*}\Phi}+\frac{\Phi^{*}i\nabla_{\mu}J^{\mu}}{2\Phi^{*}\Phi},
\end{aligned}
\label{eq:complex-matter-field-derivatives}
\end{align}
where we use the facts that $\mathrm{D}$ is $\partial$ if applied
to a scalar and $\nabla$ if applied to a vector, and $\mathrm{D}_{\mu}\left(\Phi^{*}\right)=\left(\mathrm{D}_{\mu}\Phi\right)^{*}$.
If we multiply the Euler-Lagrange equation by $\Phi$, we arrive at
\begin{align}
\begin{aligned}-m^{2}\Phi^{*}\Phi & =-\frac{J_{\mu}J^{\mu}}{4\Phi^{*}\Phi}-\frac{iJ^{\mu}}{2\Phi^{*}\Phi}\partial_{\mu}\left|\Phi\right|^{2}+\frac{\left(\mathrm{D}_{\mu}\Phi\right)^{*}\Phi iJ^{\mu}}{\Phi^{*}\Phi}+\frac{i}{2}\nabla_{\mu}J^{\mu}.\end{aligned}
\end{align}
But using (\ref{eq:DPhi-component-expressions}) we have
\begin{align}
\begin{aligned}\left\langle \mathrm{D}_{\mu}\Phi,\Phi\right\rangle _{\mathbb{C}}iJ^{\mu} & =\left(i\mathrm{Re}\left\langle \mathrm{D}_{\mu}\Phi,\Phi\right\rangle _{\mathbb{C}}-\mathrm{Im}\left\langle \mathrm{D}_{\mu}\Phi,\Phi\right\rangle _{\mathbb{C}}\right)J^{\mu}\\
 & =\left(\frac{i}{2}\partial_{\mu}\left|\Phi\right|^{2}+\frac{1}{2}J_{\mu}\right)J^{\mu},
\end{aligned}
\end{align}
so that the Euler-Lagrange equation multiplied by $\Phi$ is
\begin{align}
\begin{aligned}-m^{2}\Phi^{*}\Phi & =\frac{J_{\mu}J^{\mu}}{4\Phi^{*}\Phi}+\frac{i}{2}\nabla_{\mu}J^{\mu},\end{aligned}
\label{eq:complex-EL-eq-Phi}
\end{align}
whose real and imaginary parts yield
\begin{align}
\begin{aligned}-m^{2} & =\left\langle \mathrm{D}_{\mu}^{\varangle}\vec{\Phi},\mathrm{D}_{\mu}^{\varangle}\vec{\Phi}\right\rangle ,\\
0 & =\nabla_{\mu}J^{\mu}.
\end{aligned}
\label{eq:complex-matter-field-EOM}
\end{align}
As we see below, the second equation is redundant. Differentiating
instead by $\Phi^{*}$ yields a canonical momentum of
\begin{align}
\begin{aligned}p_{\Phi^{*}}^{\mu} & =-\frac{\Phi iJ^{\mu}}{2\Phi^{*}\Phi},\end{aligned}
\end{align}
with an Euler-Lagrange equation multiplied by $\Phi^{*}$ of
\begin{align}
\begin{aligned}-m^{2}\Phi^{*}\Phi & =\frac{J_{\mu}J^{\mu}}{4\Phi^{*}\Phi}-\frac{i}{2}\nabla_{\mu}J^{\mu},\end{aligned}
\label{eq:complex-EL-eq-PhiStar}
\end{align}
whose second term we note has a reversed sign from (\ref{eq:complex-EL-eq-Phi}).
Again, we obtain the same results if we use the equivalent Euler-Lagrange
expression $\partial L/\partial\Phi=\partial_{\mu}\left(\partial L/\partial\left(\partial_{\mu}\Phi\right)\right)$
in flat spacetime.

With the matter field EOM confirmed and a complex canonical momentum
in hand, we may apply Noether's theorem in complex form. The MFEM
Lagrangian is invariant under global infinitesimal gauge transformations,
which in complex notation transform the matter field according to
\begin{align}
\begin{aligned}\Phi & \rightarrow e^{iq\varepsilon}\Phi\\
 & =\Phi+iq\varepsilon\Phi,
\end{aligned}
\end{align}
yielding a corresponding Noether current of 
\begin{equation}
\begin{aligned}iq\Phi p_{\Phi}^{\mu} & =-\frac{1}{2}J_{q}^{\mu}\\
\Rightarrow\nabla_{\mu}J_{q}^{\mu} & =0.
\end{aligned}
\end{equation}
Thus the four-current is divergenceless if the matter field EOM are
satisfied, even if the gauge potential EOM are not; this is commonly
described in flat spacetime as ``global gauge invariance results
in conservation of charge.'' This result implies that the second
EOM in (\ref{eq:complex-matter-field-EOM}) is redundant, since it
will be true for any matter field for which the action vanishes upon
its variation.

\section{\label{sec:From-QED-to-MFEM}From QED to classical electromagnetism}

In this section we arrive at MFEM as a limit of QED by setting up
a quantum state configuration that corresponds to a classical four-current,
i.e. a configuration similar to the ``in'' state when defining the
scattering matrix, except assumed to hold locally for all spacetime.
Explicitly, this configuration will consist of spatially separated
electron wave packets interacting only via an electromagnetic field
which is too weak for pair production or bound states. Our treatment
is for spin up electrons, but spin down electrons and positrons of
both spins may be treated similarly. 

We begin with a summary of the relevant aspects of QED to fix notation
and conventions. Our presentation is a bit idiosyncratic but allows
for calculations without the usual clutter of integrals and sums.

\subsection{\label{subsec:The-QED-Lagrangian}The QED Lagrangian}

Quantum electrodynamics (in flat spacetime) is defined by quantizing
the classical field theory of the Dirac Lagrangian (minimally coupled,
using natural units and the mostly pluses metric signature)
\begin{align}
L_{\mathrm{DIRAC}} & \equiv-\mathrm{Re}\left(\overline{\psi}\gamma^{\mu}\mathrm{D}_{\mu}\psi\right)-m\overline{\psi}\psi-\frac{1}{4}F_{\mu\nu}F^{\mu\nu},\label{eq:dirac-lagrangian}
\end{align}
where in the first term we only take the negative real part, $\psi$
is a complex four-component spinor matter field, $\gamma^{\mu}$ are
the Dirac matrices, and we follow Weinberg\cite{Weinberg} in defining
$\overline{\psi}\equiv\psi^{\dagger}i\gamma^{0}$ as the Dirac adjoint
of the matter field, where $\psi^{\dagger}$ is the Hermitian conjugate.
The Dirac matrices are a complex matrix representation of an arbitrary
constant orthonormal spacetime dual frame, with matrix multiplication
the action of Clifford multiplication. 

Expanding the first and last terms of the Lagrangian, we find the
EOM from varying the gauge potential are 
\begin{equation}
\begin{aligned}q\overline{\psi}i\gamma^{\nu}\psi & =\nabla_{\mu}F^{\nu\mu},\end{aligned}
\label{eq:dirac-gp-eom}
\end{equation}
allowing us to identify the real divergenceless matter four-current
\begin{equation}
\begin{aligned}J^{\nu} & \equiv\overline{\psi}i\gamma^{\nu}\psi,\end{aligned}
\label{eq:dirac-current}
\end{equation}
in terms of which (\ref{eq:dirac-gp-eom}) is Maxwell's equations. 

We note that this expression determines the components of $J$, so
that for example if in our inertial frame the only non-zero component
is $J^{0}$, then we may write $J^{\nu}=\overline{\psi}i\gamma^{0}\psi U^{\nu}$,
where $U$ is the unit vector in the $x^{0}$ direction; this expression
then holds in any inertial frame, since under a Lorentz transformation
the spinor $\psi$ is multiplied by the matrix representation of this
transformation to make it so. We also note that the particle density
associated with $J$ may be shown to be $\psi^{\dagger}\psi$, which
is positive, and this definition allows us to write the Lagrangian
as 
\begin{align}
L_{\mathrm{DIRAC}} & =-\mathrm{Re}\left(\overline{\psi}\gamma^{\mu}\partial_{\mu}\psi\right)+qJ^{\mu}A_{\mu}-m\overline{\psi}\psi-\frac{1}{4}F_{\mu\nu}F^{\mu\nu},
\end{align}
which omitting the terms in $\psi$ is identical to the quasi-gauge
EM Lagrangian. 

Varying the matter field results in the Dirac equation 
\begin{align}
\gamma^{\mu}\mathrm{D}_{\mu}\psi & =-m\psi.
\end{align}
The Dirac operator $\gamma^{\mu}\mathrm{D}_{\mu}$ is the ``square
root'' of the Laplacian, i.e. generalized to the gauge covariant
derivative we have
\begin{equation}
\begin{aligned}\left(\gamma^{\mu}\mathrm{D}_{\mu}\right)\left(\gamma^{\nu}\mathrm{D}_{\nu}\right) & =\gamma^{\mu}\gamma^{\nu}\mathrm{D}_{\mu}\mathrm{D}_{\nu}=\eta^{\mu\nu}\mathrm{D}_{\mu}\mathrm{D}_{\nu}=\mathrm{D}^{\mu}\mathrm{D}_{\mu},\end{aligned}
\end{equation}
using the properties of the Dirac matrices $\gamma^{\mu}$ (see \cite{Marsh}).
Thus the components of $\psi$ satisfy
\begin{equation}
\begin{aligned}0 & =\left(\gamma^{\mu}\mathrm{D}_{\mu}+m\right)\psi\\
\Rightarrow0 & =\left(\gamma^{\mu}\mathrm{D}_{\mu}-m\right)\left(\gamma^{\mu}\mathrm{D}_{\mu}+m\right)\psi\\
 & =\left(\mathrm{D}^{\mu}\mathrm{D}_{\mu}-m^{2}\right)\psi,
\end{aligned}
\end{equation}
the Klein-Gordon equation. 

Varying the frame results in the on-shell Hilbert SEM tensor
\begin{align}
T_{\mathrm{DIRAC}}^{\mu\nu} & =\frac{1}{2}\mathrm{Re}\left(\overline{\psi}\gamma^{\mu}\mathrm{D}^{\nu}\psi+\overline{\psi}\gamma^{\nu}\mathbf{\mathrm{D}}^{\mu}\psi\right)+T_{\mathrm{EM}}^{\mu\nu}.\label{eq:dirac-sem-tensor}
\end{align}

\subsection{Plane wave solutions}

The canonically quantized Dirac field is based on the free quantum
Dirac equation 
\begin{equation}
\gamma^{\mu}\partial_{\mu}\hat{\psi}=-m\hat{\psi},
\end{equation}
where $\hat{\psi}$ is now the time-dependent Dirac spinor operator
in the Heisenberg picture. 

This equation has plane wave solutions, which we express using the
mostly pluses chiral basis from \cite{Weinberg}, in which the Dirac
matrices are

\begin{equation}
\gamma^{0}\equiv-i\begin{pmatrix}0 & I\\
I & 0
\end{pmatrix},\;\gamma^{i}\equiv-i\begin{pmatrix}0 & \sigma_{i}\\
-\sigma_{i} & 0
\end{pmatrix},
\end{equation}
where $\sigma_{i}$ are the Pauli matrices. With this choice we may
define an (operator-valued) spin up electron plane wave solution of
four-momentum $P=mU$ by aligning $\gamma^{0}$ and the coordinate
$x^{0}$ with $U$, so that $e^{iP_{\mu}x^{\mu}}=e^{-imx^{0}}$ and
we write

\begin{align}
\begin{aligned}\hat{\psi}_{P} & \equiv\frac{1}{\sqrt{2}}\begin{pmatrix}1\\
0\\
1\\
0
\end{pmatrix}\hat{a}_{P}e^{iP_{\mu}x^{\mu}}\\
 & \equiv u\hat{a}_{P}e^{-imx^{0}},
\end{aligned}
\end{align}
where $\hat{a}_{P}$ is postulated to be the annihilation operator
for a spin up single electron state $\ket{P}$ of four-momentum $P$.
The spinor $u$ satisfies 
\begin{equation}
\begin{aligned}u^{\dagger}u & =1,\\
\gamma^{0}u & =-iu,\\
u^{\dagger}\gamma^{0}\gamma^{\nu\neq0}u & =0,
\end{aligned}
\label{eq:u properties}
\end{equation}
so that we may verify that the free Dirac equation is satisfied:

\begin{align}
\begin{aligned}\gamma^{\mu}\partial_{\mu}\hat{\psi}_{P} & =\gamma^{0}u\partial_{0}\hat{a}_{P}e^{-imx^{0}}\\
 & =-um\hat{a}_{P}e^{-imx^{0}}\\
 & =-m\hat{\psi}_{P}
\end{aligned}
\end{align}
Keeping $\gamma^{0}$ constant, a spin up electron plane wave solution
of arbitrary four-momentum $K$ may be written 
\begin{equation}
\hat{\psi}_{K}=\Lambda_{K}u\hat{a}_{K}e^{iK_{\mu}x^{\mu}},
\end{equation}
where $\Lambda_{K}$ is the matrix representation of the Lorentz transformation
which aligns $\gamma^{0}$ with $K$. The general free solution $\hat{\psi}$
is then a (complex) linear combination (integral) of such plane wave
solutions, along with similar solutions for spin down electrons and
positrons of both spins, and if $\ket{0}$ is the ground state of
the free theory, then we have
\begin{align}
\braxket{0}{\hat{\psi}}{P} & =ue^{-imx^{0}}.
\end{align}

In the interacting theory, we choose inertial coordinates and define
an (operator-valued) plane wave solution for a spin up electron of
four-momentum $P=mU=\left(E_{\mathbf{p}},\mathbf{p}\right)$ at time
$t_{0}$ to be
\begin{equation}
\hat{\psi}_{\mathbf{p}}=\Lambda_{\mathbf{p}}u\hat{a}_{\mathbf{p}}\left(t_{0}\right)e^{i\mathbf{p}\cdot\mathbf{x}},
\end{equation}
where $\hat{a}_{\mathbf{p}}\left(t_{0}\right)$ is now postulated
to be the annihilation operator for a spin up single electron state
$\ket{\mathbf{p}}_{t_{0}}$ of momentum $\mathbf{p}$ at time $t_{0}$.
The general solution at time $t_{0}$ is again a linear combination
of such plane wave solutions, so that if $\ket{\Omega}$ is the ground
state of the interacting theory and $\gamma^{0}$ and $t=x^{0}$ are
aligned with $U$, we have
\begin{align}
\braxket{\Omega}{\hat{\psi}}{\mathbf{p}}_{t_{0}} & =ue^{-imt_{0}}.
\end{align}

\subsection{Electron packets}

We may now construct a spin up electron packet. We define the quantum
state $\ket{\phi_{\mathbf{p}}}_{t_{0}}$ to be an integral of electron
states $\ket{\mathbf{k}}_{t_{0}}$ with momenta clustered around $\mathbf{p}$
such that the spinor-valued wave packet 
\begin{align}
\begin{aligned}\phi_{\mathbf{p}}\left(\mathbf{x},t_{0}\right) & \equiv\braxket{\Omega}{\hat{\psi}}{\phi_{\mathbf{p}}}_{t_{0}}\\
 & \sim\left|\phi_{\mathbf{p}}\left(\mathbf{x},t_{0}\right)\right|ue^{-imt_{0}}
\end{aligned}
\end{align}
is smooth and normalized, i.e. its modulus is close to a smooth envelope
whose square integrated over space is unity. Propagation in time is
defined by 
\begin{align}
\begin{aligned}\gamma^{\mu}\mathrm{D}_{\mu}\phi_{\mathbf{p}}\left(\mathbf{x},t\right) & =\braxket{\Omega}{\gamma^{\mu}\mathrm{D}_{\mu}\hat{\psi}}{\phi_{\mathbf{p}}}_{t_{0}}\\
 & =-\braxket{\Omega}{m\hat{\psi}}{\phi_{\mathbf{p}}}_{t_{0}}\\
 & =-m\phi_{\mathbf{p}}\left(\mathbf{x},t_{0}\right),
\end{aligned}
\end{align}
i.e. $\phi_{\mathbf{p}}\left(\mathbf{x},t\right)$ satisfies the Dirac
equation. 

It is important to note that a Gaussian wave packet at $t_{0}$ will
not remain so as it propagates in time, even in the free theory due
to the Lorentz  factors in the $\ket{\mathbf{k}}_{t_{0}}$. For fermionic
ladder operators we are also constrained to only have one plane wave
for each four-momentum value; however, if the packet is in an unbound
state, this has no practical effect since $\mathbf{k}$ can take a
continuum of values.

Recalling (\ref{eq:u properties}), the matter four-current operator
for $\hat{\psi}_{\mathbf{p}}$ is
\begin{equation}
\begin{aligned}\hat{J_{P}}^{\nu} & =\overline{\hat{\psi}}_{P}i\gamma^{\nu}\hat{\psi}_{P}\\
 & =-u^{\dagger}\hat{a}_{P}^{\dagger}e^{-iP_{\mu}x^{\mu}}\gamma^{0}\gamma^{\nu}u\hat{a}_{P}e^{iP_{\mu}x^{\mu}}\\
\Rightarrow\hat{J}_{\mathbf{p}}^{0}\left(t_{0}\right) & =-\hat{a}_{\mathbf{p}}^{\dagger}\left(t_{0}\right)e^{-imt_{0}}\gamma^{0}\gamma^{0}\hat{a}_{\mathbf{p}}\left(t_{0}\right)e^{imt_{0}}\\
 & =\hat{a}_{\mathbf{p}}^{\dagger}\left(t_{0}\right)\hat{a}_{\mathbf{p}}\left(t_{0}\right),
\end{aligned}
\end{equation}
the number operator, and the other components vanish due to our alignment
of $\gamma^{0}$. When computing the matter four-current, the term
for positrons is reversed to also give the number operator, so that
at time $t_{0}$ we have
\begin{equation}
\begin{aligned}\braxket{\phi_{\mathbf{p}}}{\hat{J}^{0}}{\phi_{\mathbf{p}}}_{t_{0}} & \sim\left|\phi_{\mathbf{p}}\left(\mathbf{x},t_{0}\right)\right|^{2}\braket{\phi_{\mathbf{p}}}{\phi_{\mathbf{p}}}_{t_{0}}\\
 & \sim\overline{\phi}_{\mathbf{p}}\left(\mathbf{x},t_{0}\right)i\gamma^{0}\phi_{\mathbf{p}}\left(\mathbf{x},t_{0}\right)\braket{\phi_{\mathbf{p}}}{\phi_{\mathbf{p}}}_{t_{0}}.
\end{aligned}
\end{equation}
For arbitrary $\gamma^{0}$ we then have 
\begin{equation}
\begin{aligned}\braxket{\phi_{\mathbf{p}}}{\hat{J}^{\nu}}{\phi_{\mathbf{p}}}_{t_{0}} & \sim\overline{\phi}_{\mathbf{p}}\left(\mathbf{x},t_{0}\right)i\gamma^{\nu}\phi_{\mathbf{p}}\left(\mathbf{x},t_{0}\right)\braket{\phi_{\mathbf{p}}}{\phi_{\mathbf{p}}}_{t_{0}}.\end{aligned}
\label{eq:packet-current}
\end{equation}

\subsection{Classical four-current configuration}

We now define a quantum state configuration that corresponds to a
classical continuous four-current. We make the following assumptions:
\begin{enumerate}
\item \textbf{Durable localized packets}: The quantum state at any time
corresponds to some number of non-overlapping packets $\sum_{n}\phi_{\mathbf{p}_{n}}$,
each of whose four-momenta changes slowly from packet to packet. The
state evolves in time such that the packets remain smooth, localized,
and separated (a nontrivial assumption per the previous section),
and without pair production. 
\item \textbf{Smooth packet distribution}: Spacetime can be split up into
cells, each of which contains an integral number of packets with approximately
equal four-momenta $P=mU$, such that the change in the number of
packets and the change in momentum per unit of space are both much
smaller than the change in the phase of the packets per unit time. 
\item \textbf{Classical gauge potential}: The electromagnetic state is dominated
at larger length scales by the classical solution, i.e. the stationary
phase approximation holds, so that the classical gauge potential equation
of motion also holds for real-valued $A_{\mu}$.
\end{enumerate}
Within any cell we may align $t=x^{0}$ and $\gamma^{0}$ with $U$,
and choose a time $t_{0}$ near the center of a cell. The first two
assumptions mean that for each packet in the cell we may make the
approximation 

\begin{align}
\begin{aligned}\phi_{\mathbf{p}}\left(\mathbf{x},t\right) & \sim\left|\phi_{\mathbf{p}}\left(\mathbf{x},t_{0}\right)\right|ue^{-im\left(t-t_{0}\right)}.\end{aligned}
\end{align}
By choosing the Weyl gauge ($A_{0}=0$) in the cell, we therefore
can neglect any change in $\phi_{\mathbf{p}}$ except in the $U$
direction, since 

\begin{align}
\begin{aligned}\gamma^{\mu}\mathrm{D}_{\mu}\phi_{\mathbf{p}} & \sim-im\gamma^{0}\left|\phi_{\mathbf{p}}\left(\mathbf{x},t_{0}\right)\right|ue^{-im\left(t-t_{0}\right)}+\gamma^{j}\mathrm{D}_{j}\phi_{\mathbf{p}}\\
 & =-m\phi_{\mathbf{p}}+\gamma^{j}\mathrm{D}_{j}\phi_{\mathbf{p}}\\
\Rightarrow\gamma^{j}\mathrm{D}_{j}\phi_{\mathbf{p}} & \sim0.
\end{aligned}
\end{align}
The Dirac equation approximately satisfied by each $\phi_{\mathbf{p}}$
is then 
\begin{equation}
\begin{aligned}\gamma^{0}\mathrm{D}_{U}\phi_{\mathbf{p}} & =-m\phi_{\mathbf{p}},\end{aligned}
\label{eq:Dirac-packet-eq}
\end{equation}
which we will call the Dirac packet equation. Note that $U$ will
be different for each cell.

Again using the first two assumptions, we may then define a spinor-valued
field $\phi$ on all of spacetime by ``smearing'' the $\phi_{\mathbf{p}_{n}}$
across cells; more precisely, we again choose inertial coordinates
and Dirac matrices which align $x^{0}$ and $\gamma^{0}$ with $U$
and in a space-like hyperplane across each cell $s\equiv\left(\mathbf{x},t_{0}\right)$
define 
\begin{equation}
\begin{aligned}\phi_{s} & \equiv\frac{1}{\sqrt{2V_{s}}}\int_{V_{s}}\sum_{n}\phi_{\mathbf{p}_{n}}\left(\mathbf{x},t_{0}\right)\mathrm{d}^{3}x\\
 & \sim\frac{1}{\sqrt{2V_{s}}}ue^{-imt_{0}}\sum_{n}\sqrt{2}\int_{V_{s}}\left|\phi_{\mathbf{p}_{n}}\left(\mathbf{x},t_{0}\right)\right|^{2}\mathrm{d}^{3}x\\
 & =\sqrt{\rho_{0}}ue^{-imt_{0}},
\end{aligned}
\end{equation}
where we assume a Gaussian envelope and at a given time
\begin{equation}
\rho_{0}\equiv\frac{N_{s}}{V_{s}}
\end{equation}
is the number of wave packets in the space-like hyperplane of the
cell divided by the volume of the hyperplane, i.e. it is the matter
rest density. We then smoothly interpolate $\mathbf{p}$ and $\rho_{0}$
between cells to arrive at a globally defined $\phi$. At any point
we therefore have 
\[
\begin{aligned}\overline{\phi}i\gamma^{0}\phi & =\overline{\phi}\phi\\
 & =\rho_{0},
\end{aligned}
\]
and the Dirac packet equation, which is only dependent upon the phase,
remains valid at any point as
\begin{align}
\begin{aligned}\gamma^{0}\mathrm{D}_{U}\phi & =-m\phi.\end{aligned}
\end{align}
\begin{figure}[H]
\noindent \begin{centering}
\includegraphics[scale=1.14]{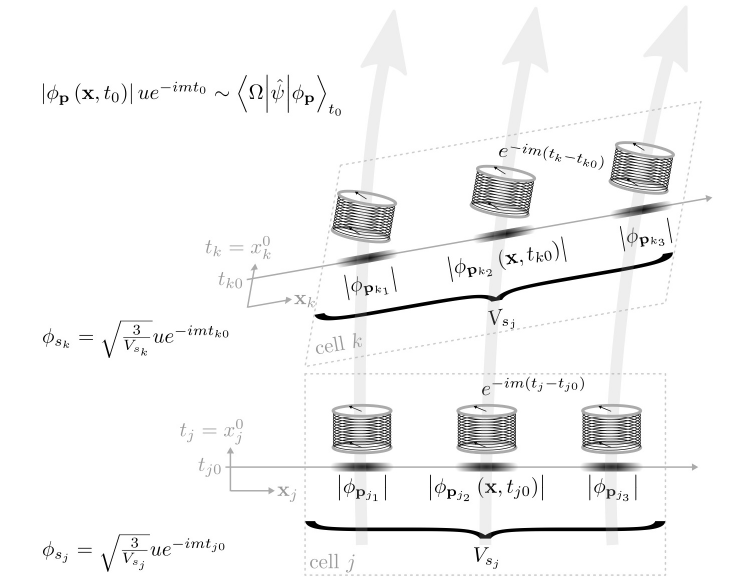}
\par\end{centering}
\caption{The quantum state configuration that corresponds to a classical continuous
four-current is a smooth distribution of durable localized packets.
By dividing spacetime into cells and aligning $t=x^{0}$ and $\gamma^{0}$
with $P=mU$, we may express each packet at time $t_{0}$ as $\braxket{\Omega}{\hat{\psi}}{\phi_{\mathbf{p}}}_{t_{0}}\sim\left|\phi_{\mathbf{p}}\left(\mathbf{x},t_{0}\right)\right|ue^{-imt_{0}}$
. In the figure the value of the modulus is represented by the shading
of each packet, while the packet phase is depicted as rotating counterclockwise
as $t$ increases. Assuming the packet momenta are approximately equal
within a cell lets us define a spinor-valued $\phi_{s}$ per cell,
which may be interpolated to yield a global spinor-valued $\phi$.
Since the change in each $\phi_{\mathbf{p}}$ and therefore $\phi$
is overwhelmingly due to the change in phase, the Dirac packet equation
$\gamma^{0}\mathrm{D}_{U}\phi=-m\phi$ is approximately satisfied
at any point, where $U=\partial/\partial t$. }
\end{figure}

Using the last assumption and (\ref{eq:packet-current}), each packet
approximately satisfies the equation of motion 
\begin{equation}
\begin{aligned}q\braxket{\phi_{\mathbf{p}}}{\hat{J}^{\nu}}{\phi_{\mathbf{p}}}_{t} & =\nabla_{\mu}F^{\nu\mu}\braket{\phi_{\mathbf{p}}}{\phi_{\mathbf{p}}}_{t}\\
\Rightarrow q\overline{\phi}_{\mathbf{p}}i\gamma^{\nu}\phi_{\mathbf{p}} & \sim\nabla_{\mu}F^{\nu\mu}.
\end{aligned}
\end{equation}
The construction of $\phi$ makes the field homogeneous within each
cell, and therefore so is $F$. Aligning $\gamma^{0}$ with $P$ in
a cell, we can then integrate over space to arrive at
\begin{equation}
\begin{aligned}\int_{V_{s}}\sum_{n}q\overline{\phi}_{\mathbf{p}}i\gamma^{\nu}\phi_{\mathbf{p}}\mathrm{d}^{3}x & \sim\int_{V_{s}}\nabla_{\mu}F^{\nu\mu}\mathrm{d}^{3}x\\
\Rightarrow q\int_{V_{s}}\sum_{n}\left|\phi_{\mathbf{p}_{n}}\left(\mathbf{x},t_{0}\right)\right|^{2}\mathrm{d}^{3}x & \sim\nabla_{\mu}F^{\nu\mu}\int_{V_{s}}\mathrm{d}^{3}x\\
\Rightarrow qN_{s} & =V_{s}\nabla_{\mu}F^{\nu\mu}\\
\Rightarrow q\rho_{0} & =\nabla_{\mu}F^{\nu\mu}\\
 & =q\overline{\phi}i\gamma^{\nu}\phi,
\end{aligned}
\label{eq:dirac-phi-gp-eom}
\end{equation}
i.e. $\phi$ satisfies the classical Dirac gauge potential equations
of motion, so that the continua of packets propagates according to
Maxwell's equations. Thus we see that the Dirac packet equation determines
$\phi$, with $U$ determined by Maxwell's equations.

Putting this all together, the equations of motion for the spinor-valued
field $\phi$ may be derived from the Lagrangian
\begin{align}
L_{\mathrm{DIRAC-\phi}} & \equiv-\mathrm{Re}\left(\overline{\phi}\gamma^{\mu}U_{\mu}\mathrm{D}_{U}\phi\right)-m\overline{\phi}\phi-\frac{1}{4}F_{\mu\nu}F^{\mu\nu}.\label{eq:dirac-classical-lagrangian}
\end{align}

\subsection{Spinor component equations of motion}

We now determine that the MFEM equations of motion are satisfied by
the rest frame components of the spinor-valued $\phi$. At any point
in spacetime we can align $\gamma^{0}$ with $P$ and write 

\begin{align}
\begin{aligned}\Phi & \equiv\sqrt{\rho_{0}}e^{-imt_{0}}\\
\Rightarrow\phi & =\frac{1}{\sqrt{2}}\begin{pmatrix}\Phi\\
0\\
\Phi\\
0
\end{pmatrix}\\
\Rightarrow J^{0} & =\overline{\phi}i\gamma^{0}\phi,\\
 & =\overline{\phi}\phi\\
 & =\left|\Phi\right|^{2}\\
 & =\rho_{0}.
\end{aligned}
\end{align}
This may be contrasted with the MFEM result $\rho_{0}=2m\left|\Phi\right|^{2}$,
and means that the gauge potential equations of motion (\ref{eq:dirac-phi-gp-eom})
may be written
\begin{equation}
\begin{aligned}q\left|\Phi\right|^{2}U^{\nu} & =\nabla_{\mu}F^{\nu\mu}.\end{aligned}
\label{eq:sc-gp-eom}
\end{equation}

If we now left multiply the Dirac packet equation by $\overline{\phi}$
and note that the right side is real, we have 

\begin{align}
\begin{aligned}\mathrm{Re}\left(\overline{\phi}\gamma^{0}\mathrm{D}_{U}\phi\right) & =-m\overline{\phi}\phi\\
\Rightarrow\mathrm{Re}\left(\phi^{\dagger}i\gamma^{0}\gamma^{0}\mathrm{D}_{U}\phi\right) & =\mathrm{Re}\left(-i\Phi^{*}\mathrm{D}_{U}\Phi\right)\\
 & =\mathrm{Im}\left\langle \Phi,\mathrm{D}_{U}\Phi\right\rangle _{\mathbb{C}}\\
 & =-\left|\Phi\right|^{2}m\\
\Rightarrow-m & =\frac{\mathrm{Im}\left\langle \Phi,\mathrm{D}_{U}\Phi\right\rangle _{\mathbb{C}}}{\left|\Phi\right|^{2}}\\
 & =\mathrm{D}_{U}^{\varangle}\vec{\Phi}.
\end{aligned}
\label{eq:dirac-equiv-eom}
\end{align}
Thus we see that for the spinor-valued field $\phi$ which represents
the ``smeared'' distribution of electron packets, the Dirac equation
is equivalent to the MFEM EOM for the rest frame components $\Phi$.
Also note that there only one sign of solution, but that building
a positron packet would have resulted in an opposite sign in the EOM;
thus both theories include anti-particles.

Lastly, at any point in spacetime we can again align $x^{0}$ and
$\gamma^{0}$ with $U$ and choose the Weyl gauge so that counting
only non-zero components we have 

\begin{align}
\begin{aligned}\overline{\phi}\gamma^{\mu}\mathrm{D}^{\nu}\phi & =\phi^{\dagger}i\gamma^{0}\gamma^{\mu}\eta^{00}\partial_{0}\phi\\
 & =im\phi^{\dagger}i\gamma^{0}\gamma^{\mu}\phi\\
 & =-m\phi^{\dagger}\gamma^{0}\gamma^{0}\phi\\
 & =m\left|\Phi\right|^{2}\\
\Rightarrow\overline{\phi}\gamma^{\mu}\mathrm{D}^{\nu}\phi & =m\rho_{0}U^{\mu}U^{\nu},
\end{aligned}
\end{align}
where in the third line we have used \ref{eq:u properties}. Thus
for an on-shell matter field the Hilbert SEM tensor using (\ref{eq:dirac-sem-tensor})
is
\begin{align}
\begin{aligned}T_{\mathrm{DIRAC-\phi}}^{\mu\nu} & =\frac{1}{2}\mathrm{Re}\left(\overline{\phi}\gamma^{\mu}\mathrm{D}^{\nu}\phi+\overline{\phi}\gamma^{\nu}\mathrm{D}^{\mu}\phi\right)+T_{\mathrm{EM}}^{\mu\nu}\\
 & =m\rho_{0}U^{\mu}U^{\nu}+T_{\mathrm{EM}}^{\mu\nu},
\end{aligned}
\end{align}
which includes the desired time-like dust term.

\subsection{From the QED to the MFEM Lagrangian}

The gauge potential and matter field EOM we have detailed for the
spinor component in the previous section may be extracted from the
Lagrangian

\begin{align}
\begin{aligned}L_{\mathrm{DIRAC-\Phi}} & =-\mathrm{Im}\left\langle \Phi,U^{\mu}\mathrm{D}_{\mu}\Phi\right\rangle _{\mathbb{C}}-m\Phi^{*}\Phi-\frac{1}{4}F_{\mu\nu}F^{\mu\nu}\\
 & =-\left\Vert \vec{\Phi}\right\Vert ^{2}\mathrm{D}_{U}^{\varangle}\vec{\Phi}-m\left\Vert \vec{\Phi}\right\Vert ^{2}-\frac{1}{4}F_{\mu\nu}F^{\mu\nu}
\end{aligned}
\end{align}
where $\Phi$ is now an arbitrary complex-valued field to be determined
by the equations of motion, and we have also expressed the Lagrangian
in terms of the real vector valued field $\vec{\Phi}$. This Lagrangian,
however, has an extra free variable: the time-like unit vector $U$
at each point, which is parallel to the four-current. It is determined
by the gauge potential EOM, but we would like to eliminate it to obtain
a standard gauge theory Lagrangian.

We may eliminate $U$ by taking advantage of the geometric observations
of Section \ref{subsec:Spacetime-gradient} and squaring the distinct
parts of the dynamical and mass terms in the Lagrangian. This results
in the MFEM Lagrangian:

\begin{align}
\begin{aligned}L_{\mathrm{DIRAC-\Phi}} & \overset{\mathrm{"squared"}}{\rightarrow}-\left\Vert \vec{\Phi}\right\Vert ^{2}\left(\mathrm{D}_{U}^{\varangle}\vec{\Phi}\right)^{2}-m^{2}\left\Vert \vec{\Phi}\right\Vert ^{2}-\frac{1}{4}F_{\mu\nu}F^{\mu\nu}\\
 & =-\left\Vert \vec{\Phi}\right\Vert ^{2}\left\langle \mathrm{D}_{\mu}^{\varangle}\vec{\Phi},\mathrm{D}_{\mu}^{\varangle}\vec{\Phi}\right\rangle -m^{2}\left\Vert \vec{\Phi}\right\Vert ^{2}-\frac{1}{4}F_{\mu\nu}F^{\mu\nu}\\
 & =L_{\mathrm{MFEM}}
\end{aligned}
\end{align}
Note that the resulting matter field EOM will include both the solutions
before squaring and the opposite sign solutions, thereby incorporating
the classical positron four-current quantum states we omitted. As
we saw in Section \ref{sec:Matter-field-electromagnetism}, the expression
of the four-current in terms of the field changes, but this change
combined with the altered Lagrangian yields the correct on-shell Hilbert
SEM tensor. 

\section{\label{sec:Summary-and-discussion}Summary and discussion}

\subsection{Summary of results}

With the Lagrangian for MFEM defined as
\begin{align}
\begin{aligned}L_{\mathrm{EM}} & \equiv-\left\langle \mathrm{D}_{\mu}^{\perp}\vec{\Phi},\mathrm{D}_{\mu}^{\perp}\vec{\Phi}\right\rangle -m^{2}\left\langle \vec{\Phi},\vec{\Phi}\right\rangle -\frac{1}{2}\left\langle F,F\right\rangle \\
 & =\frac{1}{4\left|\Phi\right|^{2}}\left(\Phi^{*}\mathrm{D}_{\mu}\Phi-\Phi\left(\mathrm{D}_{\mu}\Phi\right)^{*}\right)\left(\Phi^{*}\mathrm{D}^{\mu}\Phi-\Phi\left(\mathrm{D}^{\mu}\Phi\right)^{*}\right)-m^{2}\left|\Phi\right|^{2}-\frac{1}{4}F_{\mu\nu}F^{\mu\nu},
\end{aligned}
\end{align}
and the definition
\begin{equation}
\begin{aligned}J_{\mu} & \equiv2\left\Vert \vec{\Phi}\right\Vert \mathrm{D}_{\mu}^{\perp}\vec{\Phi}\\
 & =-i\left(\Phi^{*}\mathrm{D}_{\mu}\Phi-\Phi\left(\mathrm{D}_{\mu}\Phi\right)^{*}\right),
\end{aligned}
\end{equation}
the EOM corresponding to the variation of the matter field, gauge
potential and metric yield 
\begin{equation}
\begin{aligned}\mathrm{D}_{U}^{\varangle}\vec{\Phi} & =-m\\
qJ^{\nu} & =\nabla_{\mu}F^{\nu\mu}\\
T^{\mu\nu} & =\frac{1}{4\left\Vert \vec{\Phi}\right\Vert ^{2}}J^{\mu}J^{\nu}+F^{\mu\lambda}F^{\nu}{}_{\lambda}-\frac{1}{4}F^{\sigma\lambda}F_{\sigma\lambda}g^{\mu\nu},
\end{aligned}
\end{equation}
while the Noether current corresponding to the global gauge symmetry
is
\begin{equation}
\nabla_{\mu}J^{\mu}=0,
\end{equation}
which is also implied by the gauge potential EOM.

If we write $J=\rho_{0}U$, these EOM imply that 
\begin{equation}
\begin{aligned}\left\Vert \vec{\Phi}\right\Vert ^{2} & \begin{aligned}=\frac{\rho_{0}}{2m}\end{aligned}
,\end{aligned}
\end{equation}
which allows us to write the on-shell Hilbert SEM tensor as
\begin{equation}
\begin{aligned}T^{\mu\nu} & =\frac{1}{2}m\rho_{0}U^{\mu}U^{\nu}+F^{\mu\lambda}F^{\nu}{}_{\lambda}-\frac{1}{4}F^{\sigma\lambda}F_{\sigma\lambda}g^{\mu\nu},\end{aligned}
\end{equation}
whereupon taking the divergence of both sides yields the equations
of geodesic deviation 
\begin{equation}
m\rho_{0}\left(\nabla_{U}U\right)^{\mu}=F^{\mu}{}_{\sigma}J_{q}^{\sigma},
\end{equation}
which is equivalent to the Lorentz force law. 

One may define a QED quantum state configuration that corresponds
to a classical continuous four-current: a smooth distribution of durable
localized packets which follow the classical gauge potential EOM.
A spinor-valued field whose EOM may be derived from
\begin{align}
L_{\mathrm{DIRAC-\phi}} & \equiv-\mathrm{Re}\left(\overline{\phi}\gamma^{\mu}U_{\mu}\mathrm{D}_{U}\phi\right)-m\overline{\phi}\phi-\frac{1}{4}F_{\mu\nu}F^{\mu\nu}
\end{align}
is defined by smearing this state across spacetime; the complex-valued
rest frame components of this spinor then satisfy the MFEM EOM, which
may be derived from 

\begin{align}
\begin{aligned}L_{\mathrm{DIRAC-\Phi}} & =-\left(\mathrm{D}_{U}^{\varangle}\vec{\Phi}-m\right)\left\Vert \vec{\Phi}\right\Vert ^{2}-\frac{1}{4}F_{\mu\nu}F^{\mu\nu}.\end{aligned}
\end{align}
This Lagrangian includes an extra free variable $U$ which is determined
by the gauge potential EOM; squaring each of the terms in brackets
eliminates $U$ and yields the MFEM Lagrangian.

\subsection{\label{subsec:The-discrete-current}The discrete four-current }

The definition of the four-current $J=\rho_{0}U$ corresponding to
classical matter continua is a four-vector whose direction $U$ at
each point is tangent to worldline of the continua and whose ``length''
$\sqrt{-\left\langle J,J\right\rangle }=\rho_{0}$ is equal to the
particle number rest density $\rho_{0}$. The rest density is the
worldline rest frame particle number per unit volume (i.e. the infinitesimal
particle number per unit space-like hypersurface orthogonal to $U$),
which like $U$ is metric-dependent, so that $J$ is as well, and
these dependencies are such that the four-current density $\mathfrak{J}\equiv J\sqrt{-\det\left(g_{\mu\nu}\right)}$
is independent of the metric.

The curious fact that the $\rho_{0}$ defined in MFEM (as in Dirac
and Klein-Gordon theory) is \textit{not} metric dependent is here
explained by the QED approximation; $\rho_{0}$ is describing a unit
volume of matter that is discrete, not continuous. This means that
while $\rho_{0}$ depends upon the chosen units, it does not change
with an infinitesimal change of the metric, since an infinitesimal
change in the unit space-like hypersurface orthogonal to $U$ does
not alter the integral number of packets enclosed. 

Note that the various expressions for the four-current
\[
\begin{aligned}\overline{\psi}i\gamma^{\nu}\psi & \rightarrow\overline{\phi}i\gamma^{\nu}\phi\\
 & \rightarrow\Phi^{*}U^{\nu}\Phi
\end{aligned}
\]
are all metric-dependent, since they involve the frame; moreover,
their dependency remains consistent with discrete matter, since they
compensate for altered time-like lengths under metric variations to
keep $\sqrt{-J^{\mu}J^{\nu}g_{\mu\nu}}=\rho_{0}$ constant. The transition
to the MFEM Lagrangian results in a four-current which is a metric-independent
1-form, with $J^{\mu}=g^{\mu\nu}J_{\nu}$ no longer transforming like
a discrete four-current under metric variations. However, unlike the
various expressions for the rest density
\[
\begin{aligned}\psi^{\dagger}\psi & \rightarrow\phi^{\dagger}\phi\\
 & \rightarrow\Phi^{*}\Phi
\end{aligned}
\]
which are metric-independent, the rest density $-2\left\Vert \vec{\Phi}\right\Vert ^{2}\mathrm{D}_{U}^{\varangle}\vec{\Phi}$
in MFEM from (\ref{eq:off-shell-current}) is metric-dependent via
the unit vector $U$; when applying the matter field EOM in (\ref{eq:on-shell-current-via-matter-field}),
we instead arrive at the on-shell four-current $J^{\mu}=2\left|\Phi_{\mathrm{EL}}\right|^{2}mU^{\mu}$,
which regains its consistency with discrete matter. 

\subsection{Multiple matter fields}

Our matter field configuration is associated with a single four-current
defined on all of spacetime. Unlike in quantum theory, where there
is a single field and interacting particles are identified as Fourier
components of this field which may interfere, here we have a single
continuum four-current with no explicit reference to constituent particles. 

We may also consider multiple matter fields, each of which has its
own terms in the Lagrangian and is associated with a separate four-current;
each matter field is a separate section of the vector bundle, but
all these sections are associated with the same principal bundle with
a single parallel transport and curvature, and with the same spacetime
base space with a single metric. These four-currents will gravitationally
interact as relativistic dust via the spacetime metric, and also electromagnetically
interact via the electromagnetic gauge potential according to the
EOM $\sum_{i}qJ_{i}^{\nu}=\nabla_{\mu}F^{\nu\mu}$. 

Note that all matter fields are associated with the same charge per
particle number since they share a single field strength term, but
each may be associated with its own mass per particle number; thus
the mass per charge of each matter field is arbitrary.

One may also define a QED quantum state configuration that corresponds
to multiple classical continuous four-currents, as long as the packets
remain discrete, localized, and smoothly distributed. This results
in multiple spinor-valued fields whose multiple rest frame components
correspond to multiple MFEM matter fields.


\addcontentsline{toc}{section}{References}
\begin{thebibliography}{1}
\bibitem{Bleecker} D. Bleecker, \textit{Gauge Theory and Variational
Principles} (Addison-Wesley, 1981) 

\bibitem{Dirac} P. A. M. Dirac, \textit{General Theory of Relativity}
(Wiley, 1975) 

\bibitem{GockelerSchucker} M. Göckeler and T. Schücker, \textit{Differential
Geometry, Gauge Theories, and Gravity} (Cambridge University Press,
1987)

\bibitem{Marsh} A. Marsh, \textit{Mathematics for Physics: An Illustrated
Handbook} (World Scientific Publishing, 2018)

\bibitem{Weinberg} S. Weinberg, \textit{The Quantum Theory of Fields,
Vol. 1} (Cambridge University Press, 1995)

\bibitem{Weyl 1918}H. Weyl, ``Gravitation und Elektrizität,'' Sitz.
Berichte d. Preuss. Akad. d. Wissenschaften 465 (1918)

\bibitem{Weyl 1929}H. Weyl, ``Elektron und Gravitation,'' I. Z.
Physik 56, 330–352 (1929)

\bibitem{YangMills}C. N. Yang and R. Mills, ``Conservation of isotopic
spin and isotopic gauge invariance,'' Physical Review 96, 191–195
(1954)

\end{thebibliography}
\end{document}